\documentclass[12pt,preprint]{aastex}

\shorttitle{Anti-correlation between $M_{\odot}$ and 
$L_{\rmm bol}/L_{\rm Edd}$}
\shortauthors{Kawakatu et al.}

\begin{document}

\title{Anti-correlation between the mass of a supermassive black hole 
and the mass accretion rate in type I ultraluminous infrared galaxies 
and nearby QSOs}


\author{Nozomu Kawakatu\altaffilmark{1}}
\affil{National Astronomical Observatory of Japan, 2-21-1 Osawa, 
Mitaka, Tokyo 181-8588, Japan}

\author{Masatoshi Imanishi}
\affil{National Astronomical Observatory of Japan, 2-21-1, Osawa, Mitaka, Tokyo 181-8588, Japan}

\author{Tohru Nagao}
\affil{National Astronomical Observatory of Japan, 2-21-1 Osawa, 
Mitaka, Tokyo 181-8588, Japan}


\altaffiltext{1}{kawakatu@th.nao.ac.jp}

\begin{abstract}

We discovered a significant anti-correlation between the mass of 
a supermassive black hole (SMBH), $M_{\rm BH}$, and the luminosity 
ratio of infrared to active galactic nuclei (AGN) Eddington luminosity, 
$L_{\rm IR}/L_{\rm Edd}$, over four orders of magnitude for ultraluminous 
infrared galaxies with type I Seyfert nuclei (type I ULIRGs) and 
nearby QSOs. This anti-correlation ($M_{\rm BH}$ vs. $L_{\rm IR}/L_{\rm Edd}$) 
can be interpreted as the anti-correlation between the mass of a SMBH and
the rate of mass accretion onto a SMBH normalized by the AGN Eddington rate, 
$\dot{M}_{\rm BH}/\dot{M}_{\rm Edd}$. 
In other words, the mass accretion rate $\dot{M}_{\rm BH}$ is not 
proportional to that of the central BH mass. 
Thus, this anti-correlation indicates that BH growth is 
determined by the external mass supply process, and not the AGN Eddington-
limited mechanism. 
Moreover, we found an interesting tendency for type I ULIRGs 
to favor a super-Eddington accretion flow, whereas QSOs tended to show 
a sub-Eddington flow. 
On the basis of our findings, we suggest that a central SMBH grows by 
changing its mass accretion rate from super-Eddington to sub-Eddington. According to a coevolution scenario of ULIRGs and QSOs
based on the radiation drag process, it has been predicted that a
self-gravitating massive torus, whose mass is larger than a central BH,
exists in the early phase of BH growth (type I ULIRG phase) 
but not in the final phase of BH growth (QSO phase). 
At the same time, if one considers the mass accretion rate onto
a central SMBH via a turbulent viscosity, the anti-correlation ($M_{\rm
BH}$ vs. $L_{\rm IR}/L_{\rm Edd}$) is well explained by the positive 
correlation between the mass accretion rate $\dot{M}_{\rm BH}$ and 
the mass ratio of a massive torus to a SMBH.

\end{abstract}

\section{Introduction}

Ultraluminous infrared galaxies (ULIRGs) radiate QSO-like (= high luminosity AGN)
luminosities ($> 10^{12}L_{\odot}$) as infrared dust emission, 
and their space densities are similar to those of QSOs 
(e.g., Sanders \& Mirabel 1996). 
The AGN phenomenon appears in the final merging stage, 
and the percentage of AGNs increases with infrared luminosity 
reaching $30-50\%$ for $ L_{\rm IR} > 10^{12}L_{\odot}$ 
(Veilleux et al. 1999). 
The near-infrared light distributions in many ULIRGs appear 
to fit a de Vaucouleur's profile, which is representative of 
elliptical galaxies (Scoville et al. 2000; Veilleux et al. 2002). Plenty of molecular gas exists in their 
central kpc regions (e.g., Downes \& Solomon 1998; Bryant \& 
Scoville 1999; Gao \& Solomon 2004; Imanishi et al. 2006a), 
with gas mass densities comparable to stellar 
densities in elliptical galaxies. 
As for QSOs, their hosts are mostly luminous and 
well evolved early-type galaxies (e.g., McLeod \& Rieke 1995; 
Bahcall et al. 1997; McLure et al. 2000; Dunlop 
et al. 2003). 
Kormendy \& Sanders (1992) proposed that ULIRGs evolve 
into ellipticals through merger-induced dissipative collapse. 
Under this scenario, these mergers first go through a luminous 
starburst phase, then enter a dust-enshrouded AGN phase, 
and finally evolve into optically bright QSOs after 
they shed the dusty gas (Sanders et al. 1988). 
Until recently, it was believed that supermassive black holes (SMBHs) 
were a basic component of galaxies, and that the mass of a SMBH was tightly 
correlated to the mass, velocity dispersion, and luminosity of bulges 
(e.g., Kormendy \& Richstone 1995; Laor 1998; Magorian 
et al. 1998; Richstone et al. 1998; Ferrarese \& Merritt 2000; McLure \& Dunlop 2001, 
2002; Tremaine et al. 2002; Marconi \& Hunt 2003; Kawakatu \& Umemura 2004). 
Combined with the ULIRG-QSO connection, these findings imply that 
SMBH growth and starburst (progenitors of ellipticals) 
are physically connected. 
However, the evolutionary track between ULIRGs and QSOs has been 
an issue of long standing. 

Canalizo \& Stockton (2001) proposed that ULIRGs with type I Seyfert 
nuclei (hereafter type I ULIRGs) are the transitional stage between 
ULIRGs and QSOs because their host galaxies are undergoing tidal 
interactions. 
In addition, type I ULIRGs have a tendency to be advanced mergers with 
single nuclei (Veilleux et al. 2002).
Thus, type I ULIRGs have the attraction of the Rosetta Stone as far as revealing 
the physical relationship between ULIRGs and QSOs. 
Most of them show a full-width at half-maximum (FWHM) 
of the broad H$\beta$ line of 
less than $2000\,{\rm km}\,{\rm s}^{-1}$ (Moran et al. 1996; 
Zheng et al. 2002; hereafter 
Z02), 
and thus AGNs in type I ULIRGs would actually be narrow-line 
Seyfert 1 galaxies (NLS1s). 
Recent {\it Chandra} observations (Teng et al. 2005) 
discovered that two type I ULIRGs (IRAS [{\it
Infrared Astronomical Satellite}] F01572+0009 and IRAS Z11598$-$0112)
show a soft X-ray excess and a steep photon index ($\Gamma_{2-10\,{\rm kev}} > 2$), which are characteristic properties of NLS1s. 
All of these results suggest that type I ULIRGs have a tendency to 
harbor NLS1-like nuclei. Noting the similarities between NLS1s and 
type I ULIRGs, one possible interpretation is that type I ULIRGs have smaller central BHs than QSOs and that their BHs are rapidly growing (e.g., Pounds et al. 1995; Boller et al. 1996; Mineshige et al. 2000; Kawaguchi 2003; Collin \& Kawaguchi 
2004; Shemmer et al. 2006).
Moreover, Kawakatu et al. (2006; hereafter K06) found that 
type I ULIRGs have a BH mass one order of magnitude smaller despite 
having the comparable $R$-band bulge luminosity ($M_{\rm R}$) relative to 
QSOs and elliptical galaxies. 
They also showed that most type I ULIRGs are located near 
a proto-QSO phase, which is the early phase of BH growth predicted by 
a coevolution scenario of galactic bulges and SMBHs 
(Kawakatu et al. 2003; hereafter KUM03). 
These findings support a scenario in which type I ULIRGs 
are in the transition stage from ULIRGs to QSOs. 
However, little is understood about how SMBHs grow 
in the evolutionary sequence from ULIRGs to QSOs. 
To address this issue, we elucidated the relationship 
between the mass of a SMBH and the mass accretion rate normalized by the 
AGN Eddington rate in type I ULIRGs and QSOs. 
To this end, we derived the mass accretion rate normalized by the 
AGN Eddington rate from the ratio of AGN bolometric luminosity 
to AGN Eddington luminosity. Note that 
the infrared luminosity at 8$-$1000 $\mu$m (not far-infrared luminosity) 
is a good indicator of AGN activities for type I AGNs (type I 
ULIRGs and QSOs). 
We discuss the validity of this treatment later. 

This article is arranged as follows.
In $\S 2$, we describe how type I ULIRGs and QSOs 
were selected to achieve our aim. 
In $\S 3$, we discuss how we evaluated SMBH mass and 
infrared luminosity for type I ULIRGs and QSOs. 
Moreover, we comment on the origin of infrared luminosity 
in type I ULIRGs and QSOs.
In $\S 4$, we show the anti-correlation between the 
ratio of infrared to AGN Eddington luminosity and 
BH mass. Then, we discuss what scientists can 
learn from this anti-correlation. 
Finally, we devote $\S 5$ to discussions and conclusions. 
Throughout this article, we adopt the Hubble parameter $H_{0}=75 
\,{\rm km}\, {\rm s}^{-1}\,{\rm Mpc}^{-1}$ and the deceleration 
parameter $q_{0}=0.5$; we have converted results from 
published articles to this cosmology to facilitate comparisons.

\section{Sample Selection}

To accomplish our aim, we used a type I ULIRG sample and an optically selected 
QSO sample, for which the data of FWHM (broad ${\rm H}\beta$ line), 
the optical luminosity at $5100{\rm \AA}$ in the rest frame, and 
the infrared luminosity were available. 
The details of these samples are as follows: 

(1) The type I ULIRG sample was from Z02. 
This sample was compiled from ULIRGs in the QDOT 
(Queen Mary and Westfield College, Durham, 
Oxford, and Toronto) redshift survey (Lawrence et al. 1999), the 1Jy ULIRG survey (Kim \& Sanders 1998), 
and an IR QSO sample selected from the cross-correlation of the IRAS Point-Source Catalog with the ROSAT
({\it R\"{o}ntgensatelit}) All-Sky Survey Catalog (Boller et al. 1992). 
All type I ULIRGs selected by Z02 were ULIRGs with mid- to far-infrared properties from IRAS observations. 
From these samples, we selected all 23 objects for which Z02 provided data on both the width of the broad ${\rm H}\beta$ line and the luminosity at 
$5100{\rm \AA}$ in the rest frame. 
Except for IRAS F16136+6550, infrared luminosities $L_{\rm IR}=
L_{8-1000\,\mu{\rm m}}$ were greater than $10^{12}L_{\odot}$. 
All of them were at $\delta > -30\, {\rm degree}$, and they constituted 
$\sim 30\%$ of all type I ULIRGs identified 
in the largest IRAS redshift survey, thus providing a representative 
sample of all type I ULIRGs.
In addition, all type I ULIRGs were final merging (single-nucleus) 
objects (Veilleux et al. 2002), and the average redshift was
$z\approx 0.2$.

(2) The optically selected QSO sample comprised 47 Palomar Green quasars (PG QSOs) from Boroson \& Green (1992; hereafter BG92). 
We excluded PG 0157+001 (Mrk 1014) and PG 1226+023 (3C 273) 
from the PG QSO sample because they are categorized as type I ULIRGs. 
We added 11 PG QSOs observed by McLure \& Dunlope (2001; hereafter MD01). 
Our sample criteria were known data on the width of 
broad ${\rm H}\beta$ line, luminosity at $5100{\rm \AA}$ 
in the rest frame, and infrared luminosity. 
The average redshift of the selected QSO sample was around 0.2, and thus the redshift distribution of the PG QSOs was similar to that of the type I ULIRGs. 
Infrared flux for all 58 PG QSOs was taken from the IRAS Faint Source Catalog (Sanders et al. 1989), IRAS Point Source Catalog, or ISO Catalog (Haas et al. 2003). 

We summarize basic physical parameters of the type I ULIRGs 
and PG QSOs in Tables 1-3. 

\section{Estimation of Physical Parameters}
\subsection{Black Hole Masses}
To evaluate central SMBH mass, we assumed that the motion of ionized gas clouds moving around the BH is dominated by gravitational force and that the clouds within the broad-line region (BLR) are virialized (e.g., Peterson \& Wandel 1999, 2000). Thus, central SMBH mass can be expressed as $M_{\rm BH}\approx R_{\rm BLR}v^{2}/G$, where $v$ is the velocity dispersion of matter at the radius of the BLR $R_{\rm BLR}$, 
which is the distance of the emission-line clouds responding to the central continuum variation as determined from reverberation mappings. 
Then, central mass can be estimated as 

\begin{equation}
M_{\rm BH}=1.5\times 10^{5}\left(\frac{R_{\rm BLR}}{\rm lt-days}\right)
\left(\frac{v_{\rm FWHM}}{10^{3}{\rm km}\,{\rm s}^{-1}}\right)^{2}M_{\odot}.
\end{equation}

The velocity dispersion $v$ can be estimated from the FWHM of ${\rm H}\beta$ broad-line emission $v=fv_{\rm FWHM}$ by assuming the BLR gas to be in isotropic motion ($f={\sqrt{3}}/2$). 
Kaspi et al. (2000) found an empirical relationship between the size of the BLR, $R_{\rm BLR}$, and optical continuum luminosity, $\lambda L_{\lambda}(5100{\rm \AA})_{\rm rest}$, on the basis of 17 Seyfert galaxies and 17 optically selected PG QSOs. 

\begin{equation}
R_{\rm BLR}=32.9^{+2.0}_{-1.9}\left[\frac{\lambda L_{\lambda}(5100{\rm \AA})_{\rm rest}}{10^{44}{\rm erg}\,{\rm s}^{-1}}\right]^{0.70\pm 0.033} {\rm light}-{\rm days}.
\end{equation}

By combining equations (1) and (2), we obtain the following formula:
\begin{equation}
M_{\rm BH}=4.9^{+0.4}_{-0.3}\times 10^{6}\left[\frac{\lambda L_{\lambda}(5100{\rm \AA})_{\rm rest}}{10^{44}{\rm erg}\,{\rm s}^{-1}}\right]^{0.70\pm 0.033}\left(\frac{v_{\rm FWHM}}{10^{3}{\rm km}\,{\rm s}^{-1}}\right)^{2}M_{\odot}.
\end{equation}
Note that equation (2) holds not only for broad-line type I AGNs but also for NLS1s (Peterson et al. 2000). 
To evaluate the optical luminosity at $5100{\rm \AA}$ in the rest frame, $L_{\lambda}(5100{\rm \AA})_{\rm rest}$, we used the formula $L_{\lambda}(5100{\rm \AA})_{\rm rest}=4\pi d_{\rm L}^{2}(1+z)F_{\lambda}(5100(1+z){\rm \AA})_{\rm obs}$, where $d_{\rm L}$ is the luminosity distance. 
For 23 type I ULIRGs, the FWHM (${\rm H}{\rm \beta}$) and the observed flux at $5100{\rm \AA}$, $F_{\lambda}(5100(1+z){\rm \AA})_{\rm obs}$, were given by Z02 and measured directly from their spectra. 
For 48 PG QSOs in our sample, the FWHM of ${\rm H}{\rm \beta}$ measurements were from BG92. It is reasonable to expect that the 5100${\rm \AA}$
continuum flux measured using a 15$''$ aperture (Neugebauer et
al. 1987) is dominated by a central AGN emission in the QSO sample 
(= high luminosity AGN). 
The flux densities at $5100{\rm \AA}$ were approximated by linear 
interpolation. 
For most PG QSOs, this was performed over neighboring frequency ranges in relatively 
tight SEDs, and thus this treatment is reliable. For a few objects, $L_{\lambda}(5100{\rm \AA})_{\rm rest}$ was computed by $4400{\rm \AA}$ flux density (Kellerman et al. 1989). 
Assuming a power-law continuum $F_{\nu}\propto \nu^{\alpha}$ with a 
median optical slope of $\alpha=\alpha_{\rm opt}=-0.2$ (Neugebauer et
al. 1987), we extrapolated to 
$5100{\rm \AA}$. 
For the other 11 PG QSOs, we adopted data from the MD01 database. 

Finally, the uncertainties of SMBH mass were estimated by error propagating 
using the optical luminosity at 5100${\rm \AA}$ and the 
FWHM of the H$\beta$ measurements. The mean error of SMBH mass 
was a factor of 1.3. 
In general, SMBH mass computed in this way [eq. (3)] is systematically accurate to within a factor of 3 (e.g., Wang \& Lu 2001; Marziani et al. 2003; Schemmer et al. 2004). 

\subsection{Infrared Luminosities}

We evaluated infrared luminosities using the following formula (Sanders \& Mirabel 1996) based on the flux densities from the IRAS Faint Source Catalog (Sanders et al. 1989), IRAS Point Source Catalog, and ISO Catalog (Haas et al. 2003): 

\begin{equation}
L_{\rm IR}[8-1000\mu{\rm m}]=4\pi d_{\rm L}^{2}F_{\rm IR},
\end{equation}
where $F_{\rm IR}$ is defined as 

\begin{equation}
F_{\rm IR}=1.8\times 10^{-14}(13.48f_{12}+5.16f_{\rm 25}+2.58f_{60}
+f_{\rm 100})\,{\rm W\, m}^{-2}
\end{equation}
with $f_{12}$, $f_{25}$, $f_{60}$ and $f_{100}$ being the 
IRAS or ISO flux densities at 12, 25, 60, and 100 $\mu {\rm m}$ 
in units of Jy. 

For sources with upper limits in some bands, 
we calculated upper and lower values of infrared luminosities 
in the following ways. 
The upper value was estimated by employing the upper limit in some bands 
as the actual value. The lower value was determined by adopting zero values. 
If the difference between them was very small (less than 0.2 dex, 
which would not affect our main results), we applied the upper value 
in this article. 
If not, we show both the upper and lower values. 
As for objects with upper limits in all four bands, upper limits of 
infrared luminosities only are shown in Table 1. 


\subsection{Origin of infrared luminosities in type I ULIRGs and QSOs}
In the next section ($\S 4$), we present the $M_{\rm BH}$ vs. 
$L_{\rm IR}/L_{\rm Edd}$ relationship and comment on its physical meaning. 
Before we do, however, we discuss the origin of $L_{\rm IR}$ 
in type I ULIRGs and QSOs. 

Concerning the origin of $L_{\rm IR}$ of QSOs: 
The warmer dust emission ($T\sim 200$ K) that dominates the mid-infrared 
(12 $\mu$m and 25 $\mu$m) SEDs of QSOs is accepted as being predominantly 
AGN heated (e.g., Rowan-Robinson 1995; Haas et al. 2000, 2003). 
However, the origin of the cooler dust emission ($T\sim 50$ K) 
that dominates the far-infrared (60 $\mu$m and 100 $\mu$m) SEDs is still 
under debate (e.g., Sanders et al. 1989; Rowan-Robinson 1995; 
Haas et al. 2003; Ho 2005). 
Recently, Schweitzer et al. (2006) found that the ratio of polycyclic aromatic 
hydrocarbon (PAH) luminosity \footnote{PAH emission
is a good indicator of starburst activity because emissions from PAH
molecules are excited by far-UV photons in normal starburst galaxies,
while PAH molecules can be destroyed by X-rays from AGN (Voit 1992).} to
far-infrared luminosity ($L_{\rm FIR}=\nu L_\nu (60\mu {\rm m})$) is 
$10-30\%$ in bona fide starburst galaxies. 
This indicates that a starburst contributes $10-30\%$ of $L_{\rm FIR}$.
Because $L_{\rm IR}>L_{\rm FIR}$ (by definition) and $L_{\rm IR}$ is 
significantly larger $L_{\rm FIR}$ for QSOs (e.g., Haas et al. 2003),  
its contribution is less than $10-30\%$ for $L_{\rm IR}$.  
Therefore, the contribution of AGNs is more than $70-90\%$ for $L_{\rm IR}$; 
in other words, $L_{\rm IR}$ in QSOs is dominated by AGN power.

As for type I ULIRGs, Imanishi et al. (2006b; hereafter I06b) showed that $40\% (9/23)$ of type I ULIRGs have substantially smaller ratios of 3.3 $\mu{\rm m}$ PAH to infrared than starburst-dominated galaxies, whose luminosity ratios $L_{3.3\mu{\rm m} {\rm PAH}}/L_{\rm IR({\rm SB)}}$ are $\sim 10^{-3}$ (e.g., Mouri et al. 1990; Imanishi 2002). The scatter of PAH to the infrared luminosity ratio is a factor of $2-3$ (Fischer et al. 2000). 
Note that dust extinction in the $L$-band ($2.8-4.1\, \mu{\rm m}$) is 
less than 1 mag because $A_{\rm L}\sim 0.06 A_{\rm V}$, where $A_{\rm V}$ is 
the optical extinction (Rieke \& Lebofsky 1985; Lutz et al. 1996). 
Thus, the small $L_{3.3\mu{\rm m}, {\rm PAH}}/L_{\rm IR({\rm SB})}$ suggests that the moderately obscured ($A_{\rm V} < 15$ mag) nuclear starbursts ($<$ kpc) can contribute only a small fraction ($< 30\%$) of the infrared luminosity 
(see Table 1).
In addition, most type I ULIRGs show warm far-infrared colors ($f_{25}/f_{60} > 0.2$) \footnote{Warm infrared colors often appear in normal AGNs (e.g., de Grijp et al. 1987).}, where $f_{25}/f_{60}$ is the IRAS or ISO 25 ${\mu{\rm m}}$-to-60 ${\mu{\rm m}}$ flux ratio. 
Thus, the contributions of the host starburst and the heavily obscured nuclear 
starburst ($A_{\rm V} \gg 15$ mag) should be small. 
Moreover, the hard X-ray emission of two type I ULIRGs (IRAS F01572+0009 and IRAS Z11598$-$0112) with detected PAH emission is dominated by an AGN because their hard X-ray (2$-$10 keV)-to-AGN-bolometric-luminosity ratios are substantially larger than those in typical starburst galaxies (Teng et al. 2005). 
Therefore, these results indicate that the infrared luminosity of type I ULIRGs is powered by AGN activity. 

%
%
Finally, we question whether $L_{\rm IR}$ is the best tracer of AGN 
activity in type I ULIRGs and QSOs. 
Hao et al. (2005) claimed that for most type I ULIRGs, 
starbursts play a major role in the far-infrared band, 
whereas AGNs contribute in the mid-infrared band. Thus, one may think that 
mid-infrared luminosity ($L_{\rm MIR}=\nu L_{\nu}(12\mu{\rm m})$ or 
$\nu L_{\nu}(25\mu{\rm m})$) is a better tracer of AGN activities than 
$L_{\rm IR}$. 
As shown in Fig. 1, the anti-correlation between $M_{\rm BH}$ 
and $L_{\rm i}/L_{\rm Edd}$ can be seen in all four panels (Fig. 1a-d), 
where $L_{\rm i}$ is the monochromatic luminosity with 
$i=12\, \mu{\rm m}, 25\, \mu{\rm m}, 60\, \mu{\rm m}$ 
and $100\, \mu{\rm m}$, although the slopes of these anti-correlations are 
slightly different. 
This suggests that the $M_{\rm BH}$ vs. $L_{\rm IR}/L_{\rm Edd}$ relationship 
(Fig. 2; see discussion in $\S 4$.) has the same physical meaning 
as the $M_{\rm BH}$ vs. $L_{\rm MIR}/L_{\rm Edd}$ relationship. 
Thus, the results we present in $\S 4$ do not change whether we employ 
$L_{\rm IR}$ or $L_{\rm MIR}$. 
Next we note two advantages of using $L_{\rm IR}$ instead of 
$L_{\rm MIR}$. 
One is that we can estimate both the upper and lower values of $L_{\rm IR}$, 
even with non-detection at mid-infrared bands, as most objects are detected 
at far-infrared bands. 
Thus, if we were to employ $L_{\rm IR}$, it would be possible to use a much 
larger sample (especially for type I ULIRGs; see Table 1) 
than for $L_{\rm MIR}$, which is essential in statistical 
discussions. 
The other is that $L_{\rm IR}$ is closer than $L_{\rm MIR}$ to bolometric 
luminosity, which is important when discussing mass accretion rate onto a central BH. 
Taking account of the origin of $L_{\rm IR}$ as AGNs both in type I 
ULIRGs and QSOs and the advantages of using $L_{\rm IR}$, 
$L_{\rm IR}$ is the better choice for examining the relationship between 
mass accretion rate and central BH mass for type I ULIRGs 
and QSOs.

\section{Results}

We plot the ratio of infrared to AGN Eddington luminosity, 
$L_{\rm IR}/L_{\rm Edd}$, vs. SMBH mass, 
$M_{\rm BH}$, for 23 type I ULIRGs and 58 PG QSOs in Fig. 2. 
Filled squares denote type I ULIRGs, open squares show type I ULIRGs in which PAH emissions were detected, open circles and down arrows show broad-line QSOs (BLQSOs) whose FWHM of the broad H$\beta$ line is larger than 2000 km/s, and filled circles represent narrow-line QSOs (NLQSOs) whose FWHM of the broad H$\beta$ line is less than 2,000 km/s. 

Figure 2 shows that $L_{\rm IR}/L_{\rm Edd}$ 
was well anti-correlated with $M_{\rm BH}$ over four orders of magnitude 
(Spearman's correlation of rank coefficient $r_{\rm s}=-0.791$).
The thick solid line is the best-fitting relationship for 
all samples (type I ULIRGs and PG QSOs) except for the upper limit data, 
$\log(L_{\rm IR}/L_{\rm Edd})=
-0.961(\pm 0.081)\log{M_{\rm BH}}+7.06(\pm 0.65)$ 
with $\chi^{2}=0.319$. 
This line corresponds to the $L_{\rm IR}\approx 10^{12}L_{\odot}$. 
The dashed line represents the best-
fitting relationship for BLQSOs only except for the upper limit data, 
$\log(L_{\rm IR}/L_{\rm Edd})=-0.369 (\pm 0.108)\log{M_{\rm BH}}+1.89(\pm 0.028)$ ($\chi^{2}=0.149$). We found that the slope of the BLQSO sample 
was slightly shallower than those of the other targets. 
Next we discuss the physical meaning of the anti-correlation 
($M_{\rm BH}$ vs. $L_{\rm IR}/L_{\rm Edd}$).

\subsection{Mass accretion rate onto a SMBH in type I ULIRGs and QSOs}

As shown in Fig. 2, most type I ULIRGs (19/23) had a ratio of 
infrared to AGN Eddington luminosity of 
$L_{\rm IR}/L_{\rm Edd} > 1$. The average value of $\log(L_{\rm IR}/L_{\rm Edd})$ was 0.525 (see Table 4). 
This indicates that the AGN bolometric luminosity ($L_{\rm bol}$) of 
most type I ULIRGs
is beyond that of an AGN Eddington luminosity because $L_{\rm IR}/L_{\rm Edd}=
(L_{\rm IR}/L_{\rm bol})\times(L_{\rm bol}/L_{\rm Edd})$ and 
$L_{\rm IR}/L_{\rm bol} < 1 $ (by definition).
Theoretically, BH accretion luminosity can achieve up to 
$\approx 10L_{\rm Edd}$ (e.g., Ohsuga et al. 2005) in optically thick accretion disks called $\it{slim \,disks}$ associated with super-Eddington accretion flow (e.g., Abramowicz et al. 1988; Wang et al. 1999; Mineshige et al. 2000; Kawaguchi 2003). 
Thus, we can conclude that super-Eddington accretion flow is 
a common feature of type I ULIRGs. 
To confirm this postulate, we evaluated the AGN 
bolometric luminosity from the hard X-ray luminosity for two detected 
type I ULIRGs, assuming a bolometric correction factor of $f_{2-10{\rm keV}}=30-85$ for bright AGNs (Marconi et al. 2004; Barger et al. 2005). 
Indeed, two type I ULIRGs (IRAS F01572+0009 and IRAS Z11598$-$0112) 
showed higher ratios of AGN bolometric luminosities than BLQSOs to AGN Eddington 
luminosities (e.g., McLeod et al. 1999), $-0.44 <\log(L_{\rm bol}/L_{\rm Edd}) < 0.01$ for F01572+009 and $0.18 <\log(L_{\rm bol}/L_{\rm Edd}) < 0.63$ for Z11598$-$0112 
(see Table 5 in Teng et al. 2005). 
This supports the idea that type I ULIRGs favor higher mass accretion 
rates ($\dot{M}_{\rm BH}$) than normal BLQSOs, although the sample 
was very limited. 
Note that $\log(L_{\rm bol}/L_{\rm Edd})$ derived 
from the X-ray luminosity is less than $\log(L_{\rm IR}/L_{\rm Edd})$, 
which may imply that the bolometric correction factor $f_{2-10\,{\rm keV}}$ of 
type I ULIRGs is larger than that of BLQSOs. 
Examining the physical reasons for this in the future may prove to be worthwhile. 
 
In contrast, the infrared luminosity was less than the AGN Eddington 
luminosity for $\approx 96\%$ of QSOs, with the average 
$\log(L_{\rm IR}/L_{\rm Edd})=-1.06$ (Table 4). This coincides with 
previous findings that the mass accretion rate of normal QSOs is 
of the sub-Eddington type (e.g., McLeod et al. 1999). 
Interestingly, NLQSOs have a tendency to be higher $L_{\rm IR}/
L_{\rm Edd}$ than BLQSOs, as shown in Fig. 2. 
Three NLQSOs (PG 0050+124, PG 1244+026, and PG 1402+261) with 
$L_{\rm IR}/L_{\rm Edd}\approx 1$ showed steep hard X-ray 
power-law photon index ($\Gamma_{2-10{\rm keV}}
$), which is characteristic of AGNs with super-critical accretion,
like NLS1s. 
However, the $\Gamma_{2-10{\rm kev}}$ of two NLQSOs with 
$L_{\rm IR}/L_{\rm Edd}<1$ (PG 1211+143 and PG 1440+356) was less 
than 2 (Piconcelli et al. 2005). 
These findings confirm that $L_{\rm IR}/L_{\rm Edd}$ is a good indicator of the mass accretion rate onto a central BH for high-luminosity type I AGNs (type I ULIRGs and QSOs).


Veilleux et al. (2006) claimed that five type I ULIRGs (IRAS F07598+6508, IRAS F12540+5708, IRAS F13218+0552, IRAS F15462$-$0450, and IRAS F21219$-$1757) do not require super-Eddington accretion rates (see also Genzel et al. 2001; Tacconi et al. 2002), assuming an $H$-band early-type host magnitude-BH mass relation for well evolved early type 
galaxies as described in Marconi \& Hunt (2003). 
However, K06 showed that most type I ULIRGs are not satisfied with 
the local SMBH-bulge relationship, and thus their BH mass 
would be overestimated compared to values obtained by 
using our method [eq. (3)].
We emphasize that our conclusion (i.e., that type I ULIRGs tend to show a super-Eddington accretion rate) is based on the virial theorem without adopting 
the local SMBH-bulge relationship.

\subsection{Origin of anti-correlation between $M_{\rm BH}$ and $L_{\rm IR}/L_{\rm Edd}$}

In this section, we consider the origin of the anti-correlation between 
$M_{\rm BH}$ and $L_{\rm IR}/L_{\rm Edd}$ (Fig. 2). 
Two possible interpretations exist:

\begin{enumerate} 

\item The ratio of infrared to AGN bolometric luminosity ($L_{\rm IR}/L_{\rm bol}$) is anti-proportional to the mass of a SMBH ($M_{\rm BH}$), where $L_{\rm IR}/L_{\rm bol}$ reflects the geometry of the 
obscuring torus. 

\item The ratio of the AGN bolometric to the AGN Eddington luminosity ($L_{\rm bol}/L_{\rm Edd}$) is anti-correlated to $M_{\rm BH}$, where $L_{\rm bol}/L_{\rm Edd}$ is a tracer of the mass accretion rate normalized by the AGN Eddington mass accretion rate.

\end{enumerate}

If case (1) is a main origin of the anti-correlation, the covering factor of the obscuring torus must decrease with the mass of a SMBH. 
The clouds photoionized by AGNs' hard radiation form the narrow-line region (NLR) at $\sim 10-1000$ pc distances from AGNs because AGNs in type I ULIRGs are obscured by dust in a torus-like geometry. Thus, the strong high-excitation forbidden line emissions from NLRs (e.g., $[{\rm O}{\rm III}]\lambda 5007$) are detectable for type I ULIRGs and thus evaluate the degree of NLR development, which is reflected by the equivalent width of $[{\rm O}{\rm III}]\lambda 5007$ at the rest frame EW$([{\rm O}{\rm III}]\lambda 5007)$. 
The rest frame EW$([{\rm O}{\rm III}]\lambda 5007)$ of type I ULIRGs and PG QSOs can be taken from Z02 and BG92, respectively. 
Note that we did not use the upper values of the rest frame EW$([{\rm O}{\rm III}]\lambda 5007)$ for four type I ULIRGs (see Table 2). 
If the anti-correlation shows the difference of the obscuring geometry between type I ULIRGs and QSOs, the rest frame EW$([{\rm O}{\rm III}]\lambda 5007)$ of type I ULIRGs should be smaller than that of PG QSOs because the rest frame EW$([{\rm O}{\rm III}]\lambda 5007)$ is independent of the AGN bolometric luminosity.

To elucidate the possibility of case (1), we compared the distribution of the rest frame EW$([{\rm O}{\rm III}]\lambda 5007)$ for two samples (type I ULIRGs and PG QSOs). 
Figure 3 shows the rest frame EW$([{\rm O}{\rm III}]\lambda 5007)$ histogram. Shaded bars represent the distribution of rest frame EW$[{\rm O}{\rm III}]\lambda 5007$ for type I ULIRGs. 
Unshaded bars denote the distribution of rest frame EW$[{\rm O\,{\rm III}}]\lambda 5007$ for PG QSOs. 
To investigate whether these two distributions were statistically 
different, we applied the Kolmogorov-Smirnov statistical test, which has the advantage of making no initial assumption about the distribution of data, 
with the null hypothesis that the two distributions are from the same parent sample. The Kolmogorov-Smirnov test resulted in a probability of $18.6\%$, which indicates that the null hypothesis cannot be rejected. This result does not favor case (1). 

Thus, the physical origin of the anti-correlation ($M_{\rm BH}$ vs. $L_{\rm IR}/L_{\rm Edd}$) is the result of case (2), namely $L_{\rm bol}/L_{\rm Edd}$ is anti-correlated with $M_{\rm BH}$ over four orders of magnitude. 
Bearing in mind that $L_{\rm bol}/L_{\rm Edd}$ is a positive function of 
$\dot{M}_{\rm BH}/\dot{M}_{\rm Edd}$, 
Fig. 2 shows that $\dot{M}_{\rm BH}$ is not proportional to $M_{\rm BH}$. 
The absolute value of the mass accretion rate $\dot{M}_{\rm BH}$ can be 
given by $\dot{M}_{\rm BH}\sim L_{\rm IR}/c^{2}\approx 0.1M_{\odot}{\rm yr}^{-1}(L_{\rm IR}/10^{12}L_{\odot})$ 
because of $L_{\rm bol}=\eta\dot{M}_{\rm BH}c^{2}$ and $L_{\rm IR}=\epsilon 
L_{\rm bol}$, where $\eta$ is assumed to be 0.1$-$0.42 (the energy conversion efficiency of a BH) and the covering factor of tori $\epsilon$ 
is 0.3$-$0.5 (e.g., Sanders et al. 1989). Note that 
the mass accretion rate we estimated here may be underestimated, 
as $L_{\rm bol}/L_{\rm Edd}$ becomes insensitive to $\dot{M}_{\rm BH}/
\dot{M}_{\rm Edd}$ in the super-critical accretion flows by  
the photon-trapping effect (e.g., Begelman 1978; Watarai et al. 2000; 
Ohsuga et al. 2002, 2005). 
In contrast to this line of reasoning, Begelman (2002) proposed that the maximum 
AGN bolometric luminosity can exceed AGN Eddington luminosity by 
a factor of 100 due to photon bubble instability 
in the radiation pressure-dominated accretion disks. If this is the case, 
our evaluation might be reasonable. 

Moreover, we found that the type I ULIRGs had a higher absolute mass accretion rate than QSOs because the average infrared luminosity of type I ULIRGs was larger than that of QSOs (see Table 4). 
Thus, we can conclude that the mass accretion process 
onto a SMBH is regulated by external mechanisms and are not AGN Eddington-
limited, which is also the case for type I Seyfert galaxies (Collin \& Kawaguchi 2004). 

On the basis of these findings (\S\S 4.1 and 4.2), we suggest that a central SMBH grows as its mass accretion rate changes from super-Eddington to sub-Eddington along an evolutionary track (type I ULIRG$\to$ NLQSO$\to$ BLQSO).

\section{Discussion and Conclusions}

To reveal how a SMBH grows in the evolutionary track from 
ULIRG into QSO, we investigated the relationship between 
SMBH mass and infrared luminosity, which is a good tracer 
of AGN activities for type I ULIRGs. Our main conclusions are the following:
\begin{enumerate} 

\item
We discovered the anti-correlation between the mass of a SMBH 
($M_{\rm BH}$) and the ratio of infrared to AGN Eddington 
luminosity ($L_{\rm IR}/L_{\rm Edd}$) over four orders of magnitude 
for type I ULIRGs and QSOs, $\log(L_{\rm IR}/L_{\rm Edd})=
-0.961(\pm 0.081)\log{M_{\rm BH}}+7.06(\pm 0.65)$. 
In the case of BLQSOs only, the slope in the $M_{\rm BH}$ vs. 
$L_{\rm IR}/L_{\rm Edd}$ diagram was shallower than those of the other targets, 

\item
Type I ULIRGs and QSOs have the same distribution of the rest frame
EW$([{\rm O}{\rm III}]\lambda 5007)$. 
Because the rest frame EW$([{\rm O}{\rm III}]\lambda 5007)$ mirrors the 
geometry of the obscuring torus, the anti-correlation ($M_{\rm BH}$ vs. 
$L_{\rm IR}/L_{\rm Edd}$) can be explained as the anti-correlation between the 
mass of a SMBH and the mass accretion rate onto a SMBH normalized by that of the 
AGN Eddington. That is to say, the mass accretion rate $\dot{M}_{\rm BH}$ 
is not proportional to the mass of a SMBH. 
Furthermore, type I ULIRGs with smaller BHs tend to show higher mass accretion rates than  QSOs. 
Hence, the anti-correlation indicates that the SMBH growth process is regulated by external mechanisms rather than the self-regulation of an accretion disk around a central BH.

\item 
Super-Eddington mass accretion flow is a characteristic of 
type I ULIRGs ($\approx 85\%$ in our sample) because the infrared 
luminosity is greater than the AGN Eddington luminosity. 
However, all BLQSOs showed that the ratio of 
infrared to AGN Eddington luminosity was less than unity, 
which indicates a sub-Eddington mass accretion rate. 
It is also interesting that for three NLQSOs with hard X-ray power-law 
photon index $\Gamma_{2-10\,{\rm keV}}>2$, the mass accretion rate 
$\dot{M}_{\rm BH}$ was between that of type I ULIRGs and BLQSOs.  

\item 
Based on all of these findings (1-3), we propose an 
evolutionary track (type I ULIRG$\to$ NLQSO$\to$ BLQSO) 
altering the mass accretion rate from super-Eddington 
to sub-Eddington. 

\end{enumerate}

Finally, it is worth discussing the physical reason for the anti-correlation $M_{\rm BH}$ vs. $L_{\rm IR}/L_{\rm Edd}$. 
Based on a coevolutionary model of SMBHs and spheroidal galaxies 
(KUM03; see also Granato et al. 2004) that takes into account angular momentum 
extraction via radiation drag (Umemura 2001; Kawakatu \& Umemura 2002; 
Sato et al. 2004), massive tori form in the early phase of BH growth like type I ULIRGs, which are accreted onto SMBHs only at the last {\it e}-folding time ($t_{e}=4\times 10^{7}{\rm yr}$). 
Thus, the mass ratio of a massive torus to a SMBH, 
$M_{\rm torus}/M_{\rm BH}$, would be much larger than unity for type I ULIRGs, 
whereas $M_{\rm torus}/M_{\rm BH}$ would be less than unity for QSOs. 
Typical length scales ($r_{\rm torus}$) of such tori can be estimated 
as $r_{\rm torus}/100\,{\rm pc}\sim 0.44\alpha M_{8}/v^{2}_{100}$, 
where $M_{8}$ is the BH plus the torus mass in $10^{8}M_{\odot}$ and $v_{100}$ is the velocity in 100 km/s. The factor $\alpha$ can be inferred by high resolution-observations of tori in nearby AGN (e.g., Jaffe et al. 1993; van der Marel et al. 1998; Davis et al. 2006) and turns out to be at least a few $\alpha \sim$. 
According to KUM03, the dust emission from massive tori in type I ULIRGs is cooler than that in QSOs because of the higher column density of 
massive tori. This coincides with observational data indicating 
that the average infrared color ($f_{25}/f_{60}$) of type I ULIRGs is 
cooler than that of QSOs, as shown in Table 4.

In such a massive torus, the viscosity works effectively because 
the timescale for viscous accretion is proportional to its specific 
angular momentum, which can be reduced by radiation drag. 
Thus, a massive torus is likely to be a self-gravitating viscous object 
(e.g., Umemura 2004; Kawakatu \& Umemura 2005; and references therein).
As a plausible process of mass accretion onto a central BH, 
we consider the turbulent viscous drag whose timescale 
is $t_{\rm vis}\sim r^{2}/\nu$, where $r$ is the distance from galactic 
nuclei and $\nu$ is the viscous coefficient. 
Here we adopt $\nu=R^{-1}_{\rm crit}rv_{\rm \phi}$, where $R_{\rm crit}=100-1000$ and  $v_{\phi}$ are the critical Reynolds numbers for the onset of turbulence (e.g., Duschl et al. 2000; Burkert \& Silk 2001) and the rotation velocity, respectively. 
Then, the viscous time can be given by $t_{\rm vis}=R_{\rm crit}
t_{\rm dyn}$, where the dynamical timescale should be determined by 
the central BH plus a surrounding massive torus system.
As a consequence, the mass accretion rate ($\dot{M}_{\rm BH}$) via the viscous drag is given as $\dot{M}_{\rm BH}\propto (M_{\rm torus}/M_{\rm BH})^{3/2}[1+(M_{\rm BH}/M_{\rm torus})]^{-1/2}$ [see also eq. (22) in Granato et al. 2004]. Combining their prediction (KUM03) and mass accretion process 
due to the turbulent viscous drag, the anti-correlation $M_{\rm BH}$ vs. $L_{\rm IR}/L_{\rm Edd}$ (or $L_{\rm bol}/L_{\rm Edd}$) can be explained. 
If this interpretation is correct, 
the rate of mass accretion onto a BH may depend on the mass ratio, $M_{\rm torus}/M_{\rm BH}$. In other words, the onset of super-Eddington mass accretion may be linked to the presence of a massive torus around a central BH. To test this postulate, it will be crucial to detect massive tori among type I ULIRGs with super-Eddington mass accretion flows through carbon monoxide and hydrogen cyanide molecular emission (see also Kawakatu et al. 2007). 
To resolve $\sim 100$ pc massive tori at the typical redshift 
of type I ULIRGs $z\approx 0.2$, use of the Atacama Large Millimeter Array 
instrument is essential because a resolution of $\approx 0.1^{\prime\prime}$ 
would be necessary.

Starburst activity around AGNs may also be 
a key ingredient in interpreting the anti-correlation ($M_{\rm BH}$ vs. $L_{\rm IR}/L_{\rm Edd}$), as the starburst leads to an effective mass accretion onto a central SMBH (e.g., Norman \& Scoville 1988; Umemura et al. 1997; Wada \& Norman 2002). If this is the case, 
the galaxies with super-Eddington mass accretion will have larger starburst luminosity than those with sub-Eddington mass accretion. 
Scientists will be able to check this trend by comparing the PAH luminosity in type I ULIRGs with that in PG QSOs.

Therefore, to reveal the key physics that theoretically control mass 
accretion onto a central BH, it is necessary to determine the 
mass accretion rate via turbulent viscous drag and other physical mechanisms connected to starburst phenomena from a massive torus to a central BH with sophisticated numerical simulations. We leave this challenge for the future. 

\acknowledgments
We thank an anonymous reviewer for constructive suggestions. 
N. K. thanks K. Wada for useful and stimulating discussions. 
M. I. is supported by Grants-in-Aid for Scientific Research
(16740117). 
T. N. acknowledges financial support from the Japan Society for 
the Promotion of Science (JSPS) through the JSPS Research Fellows. 
This study made use of the NASA/IPAC Extragalactic Database 
(NED) operated by the Jet Propulsion Laboratory, California 
Institute of Technology, under contract with the National Aeronautics 
and Space Administration.


\clearpage
\appendix

\section{Infrared $L$-band spectra of two type I ULIRG nuclei}
Observations of IRAS 16136+6550 and 22454$-$1744 were made
on 19 and 20 July 2006 (UT), respectively, using a IRCS near-infrared 
spectrograph \citep{kob00} attached to a Nasmyth focus of a Subaru
8.2-m telescope \citep{iye04}.  
The sky was clear and the observation at $K$, measured in images
taken before $L$-band spectroscopy, was $\sim$0$\farcs$5 in
full-width at half-maximum. A 0$\farcs$6-wide slit
and the $L$-grism were used with a 52-mas pixel scale. The
achievable spectral resolution is R $\sim$ 140 at $\lambda 
\sim$ 3.5 $\mu$m. A standard telescope nodding technique (ABBA
pattern) with a throw of 5 to 7$''$ along the slit was employed to
subtract background emission. The optical guider of the Subaru
telescope was used to monitor the telescope tracking. Exposure
time was 1.2 s, and 50 coadds were made at each nod
position. The total net on-source integration times
were 16 min for both sources.

HR 6360 (G5V, V=6.10) and HR 8544 (G2V, V=6.57) were observed as standard
stars, with an airmass difference of $<$ 0.1 for IRAS
16136+6550 and 22454$-$1744, respectively, to correct for
the transmission of the Earth's atmosphere. 
The magnitudes of HR 6360 and HR 8544 were estimated to be $L$ =
4.5 and $L$ = 5.1, respectively, based on their $V$-band (0.6
$\mu$m) magnitudes and V$-$L colors of the corresponding
stellar types \citep{tok00}. 

Standard data analysis procedures were employed using IRAF   
\footnote{IRAF is distributed by the National Optical Astronomy
Observatories operated by the Association of Universities
for Research in Astronomy, Inc., under a cooperative agreement
with the National Science Foundation.}.
Initially, frames taken with an A (or B) beam were subtracted from
frames subsequently taken with a B (or A) beam, and the resulting
subtracted frames were added and divided by a spectroscopic flat
image. Then, bad pixels and pixels hit by cosmic rays were replaced
with the interpolated values of the surrounding pixels. Finally, the
spectra of ULIRG nuclei and standard stars were extracted by
integrating signals over 0$\farcs$6 to 1$\farcs$5, depending on actual
signal profiles. Wavelength calibration was performed taking into
account the wavelength-dependent transmission of the Earth's
atmosphere. The spectra of ULIRG nuclei were divided by the
observed spectra of standard stars and multiplied by the spectra of
blackbodies with temperatures appropriate to individual standard stars
($T$ = 5700 K and 5830 K for HR 6360 and HR 8544, respectively).

Flux calibration was done based on signals of ULIRGs and
standard stars detected inside our slit spectra. 
To obtain an adequate signal-to-noise ratio in each element,
appropriate binning of spectral elements was performed, particularly
at $\lambda_{\rm obs}$ $<$ 3.3 $\mu$m and $>$3.9 $\mu$m in
the observed frame, where the scatter is higher than at 
$\lambda_{\rm obs}$ = 3.3-3.9 $\mu$m due to the Earth's
atmosphere.  

Both sources showed flux excesses at $\lambda_{\rm obs}$ = (1 +
$z$) $\times$ 3.29 $\mu$m, the wavelength at which the 3.3-$\mu$m PAH
emission feature peaks. We thus identified these features as the 3.3-$\mu$m 
PAH emission in Fig. 4. To estimate the strength of the 3.3-$\mu$m PAH
emission feature, we adopted a template spectral shape for the
Galactic star-forming regions and nearby starburst galaxies
(type-1 sources; Tokunaga et al. 1991) as we did
previously for other ULIRGs \citep{idm06,ima06}. 
The estimated 3.3-$\mu$m PAH fluxes were 
F(3.3PAH) $\sim$ 2.0 $\times$ 10$^{-14}$ and 4.5 $\times$ 10$^{-15}$ 
ergs s$^{-1}$ cm$^{-2}$ for IRAS 16136+6550 and
22454$-$1744, respectively.

\clearpage

\begin{table}[t]
\begin{center}
Table 1. Infrared Properties of Type I ULIRGs and PG QSOs\\[3 mm]
{\scriptsize
\begin{tabular}{lccccccccc}
\hline \hline
Name &  $z$ & $\log(\frac{L_{12}}{L_{\odot}})$  & $\log(\frac{L_{25}}{L_{\odot}})$ & $\log(\frac{L_{60}}{L_{\odot}})$ & $\log(\frac{L_{100}}{L_{\odot}})$ & $\log(\frac{L_{\rm IR}}{L_{\odot}})$ & $\frac{f(25)}{f(60)}$ & References \\
(1) & (2) & (3) & (4) & (5) & (6) & (7) & (8) & (9) \\
\hline 
& & & &Type I ULIRGs & & & & \\
\hline
F00275$-$2859 &0.279 & $<$11.96 & 11.96 & 12.18  & 11.99 &12.56 & 0.25(w) &1 \\
F01572+0009(Mrk 1014) & 0.163 & 
11.69 & 11.95 & 12.23 & 12.00 & 12.53 (11.7) & 0.22(w) & 1,3 \\
F02054+0835 & 0.345 & 12.49 & $<$12.29 & 12.29 & 12.62 & 12.95 & $<0.42$ & 1 \\
F02065+4705 & 0.132 & $<$11.42 & 11.33 & 11.69 & 11.81 & 12.10 & 0.18(c) & 1 \\
F04416+1215 & 0.089 & $<$11.41 & 11.37 & 11.66 & 11.52 & 12.02 & 0.21(w) & 1 \\
IR 06269$-$0543 & 0.117 & 11.62 & 11.91 & 12.05 & 11.79 & 12.39 & 0.30(w) & 1 \\
F07598+6508 & 0.148 & 11.89 & 11.88 & 12.00 & 11.79 & 12.42 ($<12.0$)
& 0.32(w) & 1,3 \\
F09427+1929 & 0.284&12.06 & $<$11.92 & 12.05 & 12.14&12.56 & $<0.30$& 1 \\
F10026+4347 & 0.178&
$<$11.74 & 11.58 & 11.68 & 11.66 & 12.19 &0.33(w)&1 \\
F11119+3257 & 0.189 &
11.91 & 11.91 & 12.19 & 11.95 & 12.54 & 0.22(w) &1 \\
Z11598$-$0112 & 0.151&
$<$12.04 & $<$11.90 & 12.17 & 12.00 & 12.56 (11.3) & $< 0.22$ &1,3\\
F12134+5459 & 0.150 & 
$<$11.38 & 11.17 & 11.59 & 11.71 & 12.01 & 0.16(c) & 1 \\
F12265+0219\,(3C 273) & 0.158 &
12.15 & 12.18 & 12.08 & 12.10 & 12.65 ($<12.0$) & 0.52(w) & 1,3 \\
F12540+5708\,(Mrk 231) & 0.042 &
11.64 & 11.98 & 12.20 & 11.94 & 12.50 (11.2) & 0.25(w) & 1,3 \\
F13342+3932 & 0.179 &
$<$11.72 & $<$11.72 & 11.99 & 11.93 & 12.37 & $<0.22$ & 1 \\
F15069+1808 & 0.171 &
$<$11.56 & 11.36 & 11.73 & 11.69 & 12.12 & 0.17(c) & 1 \\
F15462$-$0450&0.101 &
$<$11.23 & 11.46 & 11.89 & 11.68 & 12.15 (11.3) & 0.16(c) & 1,3 \\
F16136+6550 & 0.129 &
11.28 & 11.39 & 11.45 & 11.46 & 11.91 (11.1) & 0.36(w) & 1,3 \\
F18216+6419 &0.297&
12.48 & 12.43 & 12.50 & 12.51 & 13.00 & 0.36(w) & 1 \\
F20036$-$1547 & 0.193 &
$<$11.81 & $<$11.84 & 12.23 & 12.08 & 12.54 & $< 0.17$ & 1 \\
F20520$-$2329 & 0.206 &
$<$11.87 & $<$11.68 & 11.97 & 12.03 & 12.42 & $< 0.21$ & 1 \\
F21219$-$1757& 0.113 & 
11.54 & 11.56 & 11.56 & 11.38 & 12.04 (10.7) & 0.42(w) & 1,3 \\
F22454$-$1744 & 0.117 &
$<$11.67 & 11.39 & 11.46 & 11.33 & 12.02 (10.3) & 0.36(w) & 1,3 \\
\hline
& & & & PG QSOs & & & & \\
\hline 
PG 0003+158 & 0.450 & 
$<$12.05 & $<$12.11 & $<$11.62 & $<$11.85 & $<$12.47 & $<$-1.28 & 1 \\
PG 0007+106 (III Zw 2) &0.089&
11.01 & 10.91 & 10.64 & $<$11.02 & 11.05 & 0.76 & 2 \\
PG 0026+129 & 0.142 &
$<$10.69 & $<$10.71 & $<$10.16& $<$10.41 & $<$11.07 & $<$1.48 & 1 \\
PG 0043+039 & 0.384 &
$<$12.32 & $<$12.19 & $<$11.78 & $<$11.87 & $<$12.63 & $<$1.06 & 1 \\
PG 0050+124 (IZw 1) & 0.061&
11.42 & 11.40 & 11.34 & 11.23 & 11.88 & 0.48 & 1 \\
PG 0052+251 & 0.155 &
$<$11.41 & $<$11.45 & $<$10.78 & 11.12 &11.78 & $<$1.95 & 1 \\
PG 0804+761 &0.100&
11.23 & 11.14 & 10.69 & 10.28 & 11.52 & 1.14  & 2 \\
PG 0838+770 & 0.131 &
10.89 & 11.05 & 10.90 & 11.07 & 11.49 & 0.59 & 1 \\
PG 0844+349 (Ton 951) & 0.064 &
10.82 & 10.71 & 10.24 & 10.27 & 11.12 & 1.25 & 1\\
PG 0923+201 & 0.190 &
$<$11.68 & $<$11.56 & $<$11.47 & $<$11.77 & 
$<$12.14 & $<$0.5 & 1\\
PG 0953+414  &0.239&
$<$11.85 & $<$ 11.61 & $<$11.31 & $<$11.48 & $<$12.14 & $<$0.83 & 1 \\
PG 1004+130 & 0.240 &
$<$11.87 & 11.76 & 11.49 & $<$11.44 & 11.87 --- 12.21 & 0.78 & 1 \\
PG 1012+008 & 0.185 & $<$11.70 & $<$11.52 & $<$11.12 & $<$11.29 & $<$11.99 & 1.05 & 1 \\
PG 1049$-$005 & 0.357 &12.23 & 12.21 & 11.86 & $<$11.97 & 12.62 & 0.10 & 1 \\
PG 1100+772 (3C 249.1) & 0.313 &
11.41 & 11.44 & 11.23 & 10.83 & 11.82 & 0.73 & 2 \\
PG 1103$-$006 & 0.425 &
$<$12.51 & $<$12.31 & 11.86 & $<$12.11 & 
11.83 --- 12.79 & $<$1.17 & 1 \\
PG 1114+445 & 0.144&
11.38 & 11.29 & $<$11.12 & 10.82 &11.74 & 0.78 & 2 \\
PG 1116+215 (Ton 1388) & 0.177 &
11.81 & 11.58 & $<$11.27 & $<$11.16 & 12.07 & $>$0.85 & 2 \\
PG 1119+120\,(Mrk 734) & 0.049 &
10.57 & 10.62 & 10.52 & 10.44 & 11.07 & 0.51 & 1 \\
PG 1149$-$110 & 0.049 &
10.57 & 10.60 & 10.35 & 10.06 & 10.98 & 0.74 & 2 \\
PG 1202+281 & 0.165 & 
$<$11.56 & 11.34 & 10.91 & $<$11.27 & 11.56 --- 11.85 & $>$1.13 & 1 \\
\hline
\end{tabular}
}
\noindent
\end{center}
\end{table}

\begin{table}[t]
\begin{center}
Table 1. - Continued \\[3mm]
{\scriptsize
\begin{tabular}{lccccccccc}
\hline \hline
Name &  $z$ & $\log(\frac{L_{12}}{L_{\odot}})$  & $\log(\frac{L_{25}}{L_{\odot}})$ & $\log(\frac{L_{60}}{L_{\odot}})$ & $\log(\frac{L_{100}}{L_{\odot}})$ & $\log(\frac{L_{\rm IR}}{L_{\odot}})$ & $\frac{f(25)}{f(60)}$ & References \\
(1) & (2) & (3) & (4) & (5) & (6) & (7) & (8) & (9) \\
\hline \hline
PG 1229+204\,(Mrk771) & 0.064 &
10.56 & 10.60 & 10.40 & $<$10.30 & 11.20 & 1.94 & 1 \\
PG 1244+026 & 0.048 &
10.40 & 10.46 & 10.34 & 10.11 & 10.87 & 0.56 & 2 \\
PG 1259+593 & 0.472 &
$<$12.59 & $<$12.30 & $<$12.03 & $<$12.16 & $<$12.86 & $<$0.78 & 1 \\
PG 1302$-$102 & 0.286 &
$<$12.14 & $<$12.01 & $<$11.59 & $<$11.76 &12.46 & $<$1.07 
& 1 \\
PG 1307+085 & 0.155 &
$<$11.56 & $<$11.38 & $<$11.00 & $<$11.13 & 11.86 
& $<$0.94 & 1 \\
PG 1309+355 (Ton 1565) & 0.184 & 
11.44 & 11.35 & $<$11.18 & $<$11.03 & 11.65 --- 11.81 & $>$0.63 & 2 \\
PG 1322+659 &0.168&
11.44 & 11.039 & 10.83 & 11.07 & 11.47 & 0.6 & 2 \\
PG 1351+236 & 0.055 &
$<$10.61 & $<$10.32 & 10.45 & 10.29 & 
10.99 --- 11.38 & $<$0.31 & 1 \\
PG 1351+640 & 0.087 &
11.12 & 11.42 & 11.15 & 10.79 & 11.29 & 0.76 & 2 \\ 
PG 1352+183 & 0.158 &
$<$11.55 & $<$11.26 & $<$10.97 & $<$11.15 & 11.22 & $<$0.81 & 2 \\
PG 1354+213 &0.300 &
$<$12.13 & $<$11.85 & $<$ 11.60 & $<$11.73 & $<$12.41 & $<$0.73 & 1\\
PG 1402+261 & 0.164 &
11.40 & 11.32 & 11.18 & 10.97 & 11.81 & 0.52 & 1 \\
PG 1411+442 &0.089 &
11.07 & 10.90 & 10.52 & 10.34 & 11.34 & 0.99 & 1 \\
PG 1415+451 & 0.114 &
10.88 & 10.77 & 10.58 & 10.48 & 11.24 &0.65  & 2 \\
PG 1416$-$129 & 0.129 &
$<$11.38 & $<$11.28 & $<$ 10.79 & $<$10.92 & $<$11.69 & $<$1.29 & 1 \\
PG 1425+267 & 0.366 &
$<$12.27 & $<$ 11.91 & $<$11.66 & $<$11.88 & $<$12.52 & $<$0.74 & 1  \\
PG 1426+015 (Mrk 1383) & 0.086 &
11.081 & 10.90 & 10.79 & $<$ 10.56 & 11.39 & 0.54 & 2 \\
PG 1427+480 & 0.221 &
11.66 & 11.33 & 11.14 & 11.27 & 11.93 & 0.58 & 2 \\
PG 1435$-$067 & 0.129 &
11.33 & $<$11.21 & 10.75 & $<$10.92 & 11.59 & $<$0.41 & 2 \\
PG 1440+356\,(Mrk478)&0.077 &
10.99 & 10.89 & 10.96 & 10.86 & 11.46 & 0.35 & 2 \\
PG 1444+407 & 0.267 &
11.78 & 11.72 & 11.40 & 10.99 & 12.11 & 0.86 & 2 \\
PG 1501+106 (Mrk 841) & 0.036 &
10.25 & 10.53 & 10.20 & $<$9.75 & 10.79 & 0.89 & 1 \\
PG 1512+370 (4C37.43) & 0.371 &
11.87 & 11.69 & 11.43 & $<$11.32 & 12.17 & 0.77 & 1 \\
PG 1519+226 &0.137 &
11.26 & 10.94 & $<$10.75 & $<$10.88 & 11.51 & $>$0.57 & 2 \\
PG 1534+580 (Mrk 290)& 0.030 &
10.057 & 10.00 & 9.56 & 9.61 & 10.39 & 1.13 & 1 \\
PG 1545+210 (3C323.1) & 0.266 &
11.40 & 11.32 & 10.96 & 10.56 & 11.71& 0.96 & 2 \\
PG 1612+261 &0.131 &
$<$10.88 & $<$10.64 & $<$10.39 & $<$10.64 & $<$11.19 & $<$0.96 & 1 \\
PG 1613+658(Mrk876) & 0.129 &
11.28 & 11.39 & 11.45 & 11.46 & 11.91 & 0.36 & 1 \\
PG 1626+554 & 0.133 &
10.95 & 10.64 & $<$10.52 & 10.83 & 11.33 & $>$0.32 & 2\\
PG 1704+608\, (3C 351) & 0.371 &
12.01 & 12.04 & 11.92 & 11.57 & 12.46 & 0.53 & 2 \\
PG 2112+059 & 0.466 &
12.38 & 12.07 & 11.85 & $<$11.86 & 12.64 & 0.69 & 1 \\
PG 2130+099 (II Zw 136) &0.061 &
10.95 & 10.88 & 10.66 & 10.44 & 11.32 & 0.69 & 2 \\
PG 2251+113 & 0.323 &
$<$11.74 & $<$11.68 & $<$11.31 & $<$11.59 & $<$12.13 & 0.98 & 1\\
PG 2308+098 (4C 09.72)& 0.432 &
11.76 & 11.89 & $<$12.01 & $<$11.75 & 12.07 --- 12.39 & $>$0.31 & 2\\
PG 2349$-$014 & 0.173&
$<$11.69 & $<$11.55 & 11.35 & 11.16 & 11.47 --- 12.02 & $<$0.66 & 1\\
\hline 
\hline
\end{tabular}
}
\noindent
\end{center}
{\scriptsize Note.-Col. (1): Source name. Col. (2): Redshift. Cols. (3-6): $L_{12}$, $L_{25}$, $L_{60}$, and $L_{100}$ are the monochromatic luminosities ($\nu L_{\nu}$) at 12$\mu$m, 25$\mu$m, 60$\mu$m, and 100$\mu$m, respectively. 
Col. (7): Infrared luminosity and the values in parentheses are infrared luminosity due to starburst, $\log(\frac{L_{\rm IR(SB)}}{L_{\odot}})$, which was derived by $L_{3.3{\rm PAH}}/L_{\rm IR(\rm SB)}=10^{-3}$ for starburst-dominated galaxies (e.g., Mouri et al. 1990; Imanishi 2002). Polycyclic aromatic hydrocarbon luminosity was based on Imanishi et al. (2006b) and recent observations (see Appendix). 
Col. (8): Ultraluminous infrared galaxies with type I Seyfert nuclei (type I ULIRGs) with $f(25)/f(60)$$<$0.2 and $>0.2$ are classified as cool and warm (represented as C and W, respectively; Sanders et al. 1988). Col. (9): References.-(1) Sanders et al. (1989); (2) Haas et al. (2003); 
(3) Imanishi et al. (2006b).\\
}
\end{table}

\begin{table}[t]
\begin{center}
Table 2. Variable Physical Parameters for Ultraluminous Infrared Galaxies With Type I Seyfert Nuclei  \\[3mm]
{\scriptsize
\begin{tabular}{lccccc}
\hline \hline
Name & $\log(\frac{M_{\rm BH}}{M_{\odot}})$ & $\log(\frac{L_{\rm IR}}{L_{\rm Edd}})$ &${\rm EW[O\,III]}({\rm \AA})$ & References \\
(1) & (2) & (3) & (4) & (5) \\
\hline 
F00275$-$2859 & 7.27 &  0.78 & 10.7 &1 \\
F01572+0009(Mrk 1014)$^{a}$ & 7.87 &  0.15 & 45.8 & 1 \\
F02054+0835 & 7.68 &  0.76 &  9.2 & 1 \\
F02065+4705 & 6.89 &  0.69 & 13.6 & 1 \\
F04416+1215 & 6.66 &  0.85 & 28.0 & 1 \\
IR 06269$-$0543 & 6.93 & 0.95 & 85.2 & 1 \\
F07598+6508 & 8.17  & -0.26 & $<$1.1 & 1 \\
F09427+1929 & 7.50 &  0.55 & $<$1.1 & 1 \\
F10026+4347 & 7.54 & 0.14 & 3.8 & 1 \\
F11119+3257 & 7.24 & 0.79  & 38.5 & 1\\
Z11598$-$0112$^{a}$ & 6.36 & 1.69 & 10.5 & 1\\
F12134+5459 & 6.10 & 1.40 & 22.8 & 1 \\
F12265+0219\,(3C 273) & 9.00 & -0.86 & 4.9 & 1 \\
F12540+5708\,(Mrk 231) & 7.94 & 0.05 & $<$4.1 & 1\\
F13342+3932 & 6.58 & 1.28 & 38.0 & 1 \\
F15069+1808 & 6.92 & 0.71 & 47.2 & 1 \\
F15462$-$0450 & 6.31 & 1.33 & 32.2 & 1 \\
F16136+6550 & 8.70 & -1.30 & 11.7 & 1 \\
F18216+6419 & 9.13 & -0.64 & 22.9 & 1 \\
F20036$-$1547 & 7.32 & 0.71 & $<$0.3 & 1\\
F20520$-$2329 & 7.15 & 0.79 & 10.3 & 1\\
F21219$-$1757 & 7.15 & 0.38 & 11.2 & 1 \\
F22454$-$1744 & 6.32 & 1.19 & 35.8 & 1 \\
\hline
\end{tabular}
}
\noindent
\end{center}
{\scriptsize Note.-Col. (1): Source name. 
Col. (2): Black hole mass. Col. (3): $L_{\rm IR}/L_{\rm Edd}$. Col. (4): Equivalent width of $[{\rm O}\,{\rm III}]\lambda 5007{\rm \AA}$ at the rest frame. Col. (5): References (for cols. [2] and [4]). \\
References.-(1) Zheng et al. (2002).\\
The symbol $a$: Object showed a soft X-ray excess and a steep photon index 
$\Gamma_{2-10\,{\rm kev}} > 2 $ (Teng et al. 2002).

}
\end{table}

\clearpage
\begin{table}[t]
\begin{center}
Table 3. Variable Physical Parameters for PG QSOs\\[3 mm]
{\scriptsize
\begin{tabular}{lccccc}
\hline \hline
Name & $\log(\frac{M_{\rm BH}}{M_{\odot}})$ & $\log(\frac{L_{\rm IR}}{L_{\rm Edd}})$ &${\rm EW[O\,III]}({\rm \AA})$ &References \\
(1) & (2) & (3) & (4) & (5) \\
\hline 
\hline
PG 0003+158 & 9.34 & $<$-1.38 & 26 & (2\&4),2 \\
PG 0007+106 (III Zw 2) & 8.29 & $-$1.75 & 42 & (2\&3),2 \\
\underline{PG 0026+129} & 7.85 & $<$-1.29 & 29 & (2\&3),2 \\
PG 0043+039 & 9.23 & $<$-1.11 & 1 & (2\&3),2 \\
\underline{PG 0050+124\,(IZw 1)}$^{a}$ & 7.26 & 0.11 & 22 & (2\&4),2 \\
PG 0052+251 & 8.72 & $-$1.45 & 30 & (2\&3),2 \\
PG 0804+761 & 8.31 & $-$1.30 & 9 & (2\&3),2 \\
PG 0838+770 & 7.99 & $-$1.01 & 13 & (2\&3),2 \\
PG 0844+349 (Ton 951) & 7.69 & $-$1.08 & 8 & (2\&3),2 \\
PG 0923+201 & 8.95 & $<$-1.32 & 7 & 1,2\\
PG 0953+414 & 8.58 & $<$-0.95 & 18 & (2\&3),2 \\
PG 1004+130 & 9.11 & $-$1.75 --- $-$1.41 & 6 & 1,2 \\
PG 1012+008 & 7.80 & $<$-0.32 & 29 & 1,2 \\
PG 1049$-$005 & 9.14 & $-$1.03 & 55 & (2\&3),2 \\
PG 1100+772 (3C 249.1) & 9.07 & $-$1.76 & 41 & (2\&3),2 \\
PG 1103$-$006 & 9.29 &$-$1.97 --- $-$1.01 & 8 & (2\&3),2 \\
PG 1114+445 & 8.41 & $-$1.18 & 14 & (2\&3),2 \\
PG 1116+215 (Ton 1388) & 8.22 & $-$0.66 & 10 & 1,2 \\
\underline{PG 1119+120\,(Mrk 734)} & 7.20 & $-$0.64 & 19 & (2\&3),2 \\
PG 1149$-$110 & 7.57 & $-$1.10 & 33 & (2\&3),2 \\
PG 1202+281 & 8.30 & $-$1.25 --- $-$0.96 & 36 & 1,2 \\
\underline{PG 1211+143}$^{b}$ & 7.88 & $-$0.81 & 12 & (2\&3),2 \\
PG 1216+069 & 9.17 & $-$1.35 & 10 & (2\&3),2 \\
PG 1229+204\,(Mrk771) & 7.93 & $-$1.24 & 19 & (2\&3),2 \\
\underline{PG 1244+026}$^{a}$ & 6.29 &  0.07 & 17 & (2\&3),2 \\
PG 1259+593 & 8.98 & $<$-0.63 & 0 & (2\&3),2 \\
PG 1302$-$102 & 8.31 & $-$0.8 --- $-$0.36 & 9 & 1,2 \\
PG 1307+085 & 7.86 & $-$1.15 --- $-$0.51 & 32 & 1,2 \\
PG 1309+355 (Ton 1565) & 8.21 & $-$1.07 --- $-$0.91 & 19 & (2\&3),2 \\
PG 1322+659 & 8.16 & $-$1.20 & 8 & (2\&3),2 \\
PG 1351+236 & 8.31 & $-$1.83 --- $-$1.44& 12 & (2\&3),2 \\
PG 1351+640 & 8.69 & $-$1.91 & 12 & (2\&3),2 \\ 
PG 1352+183 & 8.27 & $-$1.56 & 10 & (2\&3),2 \\
PG 1354+213 & 8.54 & $<$-0.64 & 31 & (2\&3),2 \\
\underline{PG 1402+261}$^{a}$ & 7.30 & 0.0 & 1 & 1,2 \\
PG 1411+442 & 7.89 & $-$1.06 & 15 & (2\&3),2 \\
PG 1415+451 & 7.81 & $-$1.08 & 1 & (2\&3),2 \\
PG 1416$-$129 & 8.50 & $<$-1.32 & 49 & (2\&3),2 \\
PG 1425+267 & 9.78 & $<$-1.77 & 36 & (2\&3),2  \\
PG 1426+015 (Mrk 1383) & 8.75 & $-$1.87 & 11 & (2\&3),2 \\
PG 1427+480 & 8.00 & $-$0.58 & 58 & (2\&3),2 \\

\hline
\end{tabular}
}
\end{center}
\end{table}

\begin{table}[t]
\begin{center}
Table 3. - Continued \\[3mm]
{\scriptsize
\begin{tabular}{lccccc}
\hline \hline
Name & $\log(\frac{M_{\rm BH}}{M_{\odot}})$ & $\log(\frac{L_{\rm IR}}{L_{\rm Edd}})$ &${\rm EW[O\,III]}({\rm \AA})$ & References \\
(1) & (2) & (3) & (4) & (5) \\
\hline 
\hline 
PG 1435$-$067 & 8.24 & $-$1.16 & 12 & (2\&3),2 \\
\underline{PG 1440+356\,(Mrk478)}$^{b}$ & 7.30 & $-$0.35 & 10 & (2\&3),2 \\
PG 1444+407 & 8.23 & $-$0.63 & 1 & (2\&3),2 \\
PG 1501+106 (Mrk 841) & 7.88 & $-$1.60 & 64 & (2\&3),2 \\
PG 1512+370 (4C37.43) & 8.95 & $-$1.29 & 57 & (2\&3),2 \\
PG 1519+226 & 7.78 & $-$0.78 & 4 & (2\&3),2 \\
PG 1534+580 (Mrk 290) & 7.36 & $-$1.48 & 79 & (2\&3),2 \\
PG 1545+210 (3C323.1) & 8.94 & $-$1.74 & 33 & 1,2 \\
PG 1612+261 & 7.91 & $<$-1.23 & 157 & (2\&3),2 \\
PG 1613+658(Mrk876) & 8.99 & $-$1.59 & 20 & (2\&3),2 \\
PG 1626+554 & 8.36 & $-$1.54 & 9 & (2\&3),2 \\
\underline{PG 1704+608\, (3C 351)} & 7.88 & 0.07 & 27 & (2\&3),2 \\
PG 2112+059 & 9.14 &  $-$1.01 & 0 & (2\&3),2 \\
PG 2130+099 (II Zw 136) & 7.68 & $-$0.87 & 20 & (2\&3),2 \\
PG 2251+113 & 8.96 & $<$-1.34 & 19 & (2\&3),2 \\
PG 2308+098 (4C 09.72) & 9.57 & $-$2.01 --- $-$1.69 & 17 & (2\&4),2 \\
PG 2349$-$014 & 8.79 & $-$1.83 --- $-$1.28 & --- & 1,2 \\
\hline 

\end{tabular}
}
\noindent
\end{center}
{\scriptsize Note.-Col. (1): Source name. Col. (2): Black hole mass. Col. (3): $L_{\rm IR}/L_{\rm Edd}$. Col. (4): Equivalent width of $[{\rm O}\,{\rm III}]\lambda 5007{\rm \AA}$ at the rest frame. Col. (5): References (for cols. [2] and [4], respectively). \\
References.-(1) McLure \& Dunlop (2001) for the full width at half maximum [FWHM(H$\beta$)] and the $L_{\lambda}(5100{\rm \AA})_{\rm rest}$; (2) Boroson \& Green (1992) for the FWHM(H$\beta$); (3) Neugebauer et al. (1987) for the $L_{\lambda}(5100{\rm \AA})_{\rm rest}$; (4) Schmidt \& Green (1983) for the $L_{\lambda}(5100{\rm \AA})_{\rm rest}$.\\
Underlined objects are narrow-line QSOs whose FWHMs of the broad H$\beta$ lines were less than 2000 km/s.\\ 
The symbol $a$: Object showed a steep hard X-ray photon index $\Gamma_{2-10{\rm kev}} >2$. The symbol $b$: Hard X-ray photon index $\Gamma_{2-10{\rm kev}} < 2$ (Piconcelli et al. 2005). 
}
\end{table}

\begin{table}[t]
\begin{center}
Table 4.  Comparison of Several Physical Properties \\[3mm]
{\scriptsize
\begin{tabular}{lccccccc}
\hline \hline
Name &$<\log(\frac{L_{\rm IR}}{L_{\odot}})>$ & $<\log(\frac{M_{\rm BH}}{M_{\odot}})>$ & $<\log(\frac{L_{\rm IR}}{L_{\rm Edd}})>$ & $<\frac{f(25)}{f(60)}>$\\
(1) & (2) & (3) & (4) & (5) \\
\hline
Type I ULIRGs & 12.37 (0.30) &7.26 (0.84)  & 0.525 (0.74)  & 0.270 ($9.8\times 10^{-2}$) \\
QSOs & 11.56 (0.51) & 8.33 (0.68) & $-$1.06 (0.55) & 0.766 (0.33) \\
\hline
\end{tabular}
}
\noindent
\end{center}
{\scriptsize Note.-Col. (1): Name of subgroup.  Col. (2): Average infrared luminosity and its dispersion, except for objects with upper and lower limits. Col. (3): Average black hole mass and its dispersion. Col. (4): Average $L_{\rm IR}/L_{\rm Edd}$ and its dispersion, except for the with upper and lower limits. Col. (5): Average $\frac{f(25)}{f(60)}$ and its dispersion, except for objects with upper and lower limits. Type I ULIRG = ultraluminous infrared galaxies with type I Seyfert nuclei.

}
\end{table}

\clearpage
\begin{figure}
\begin{center}
\includegraphics[width=6cm]{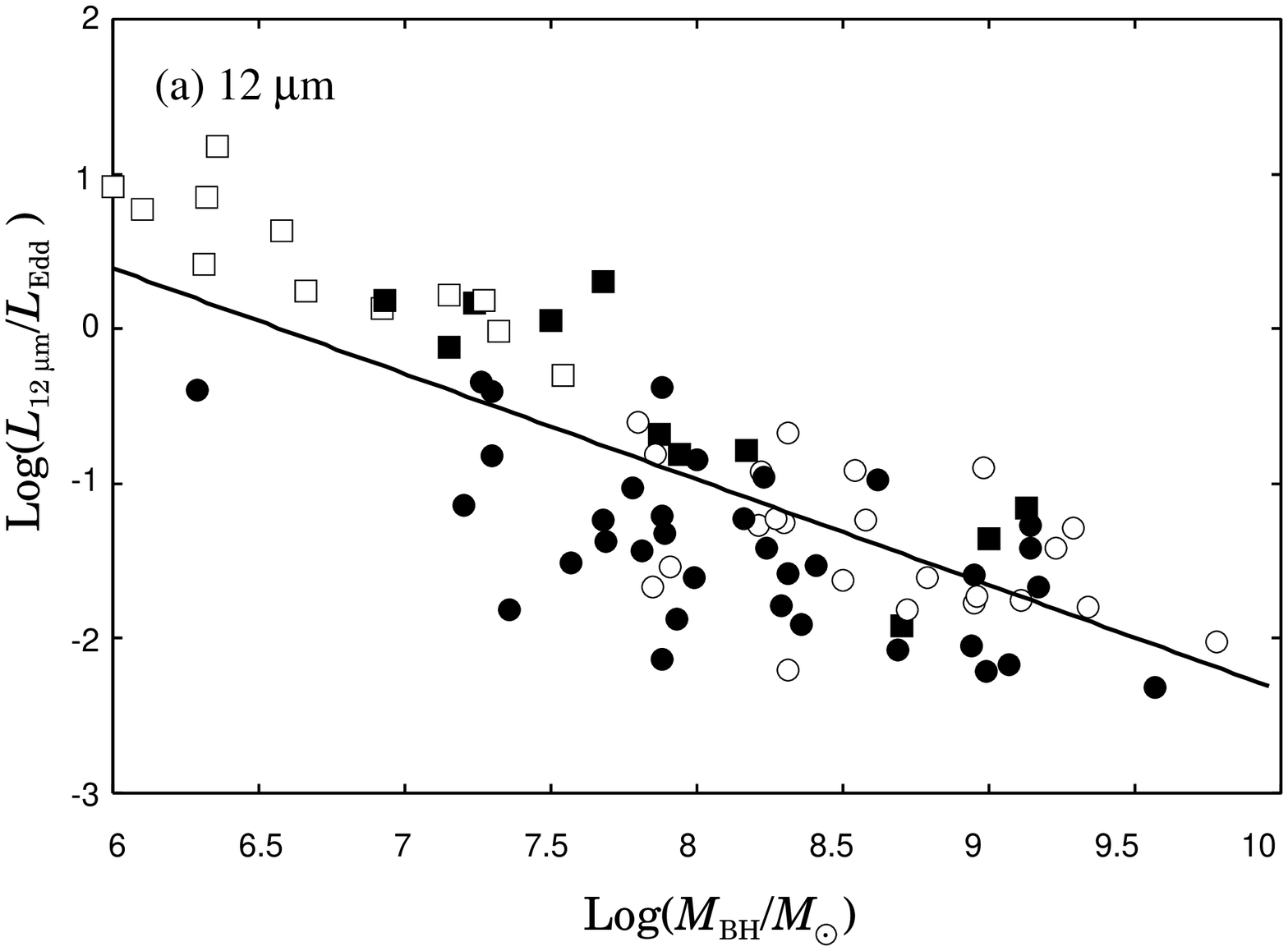}
\includegraphics[width=6cm]{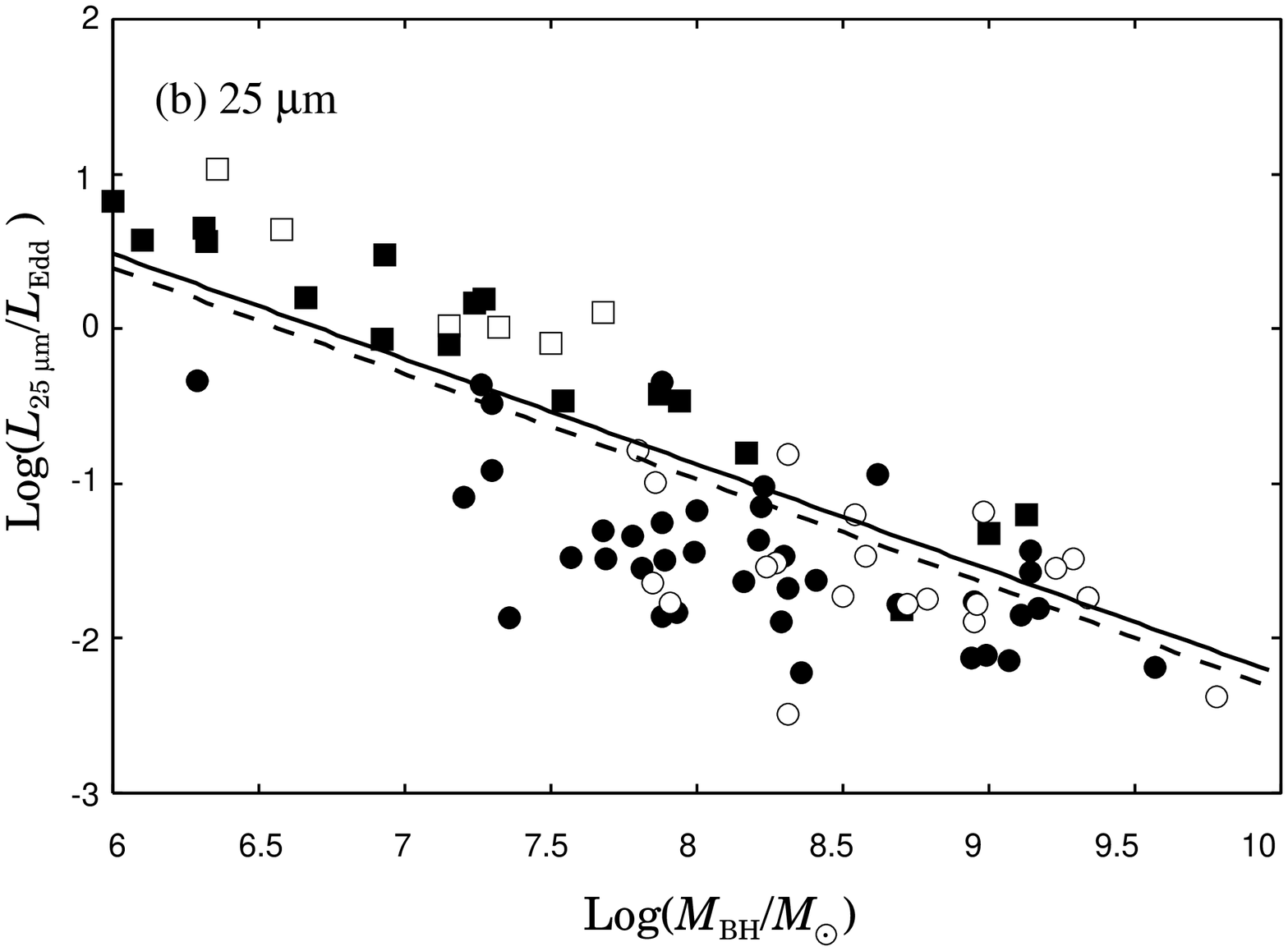} \\
\includegraphics[width=6cm]{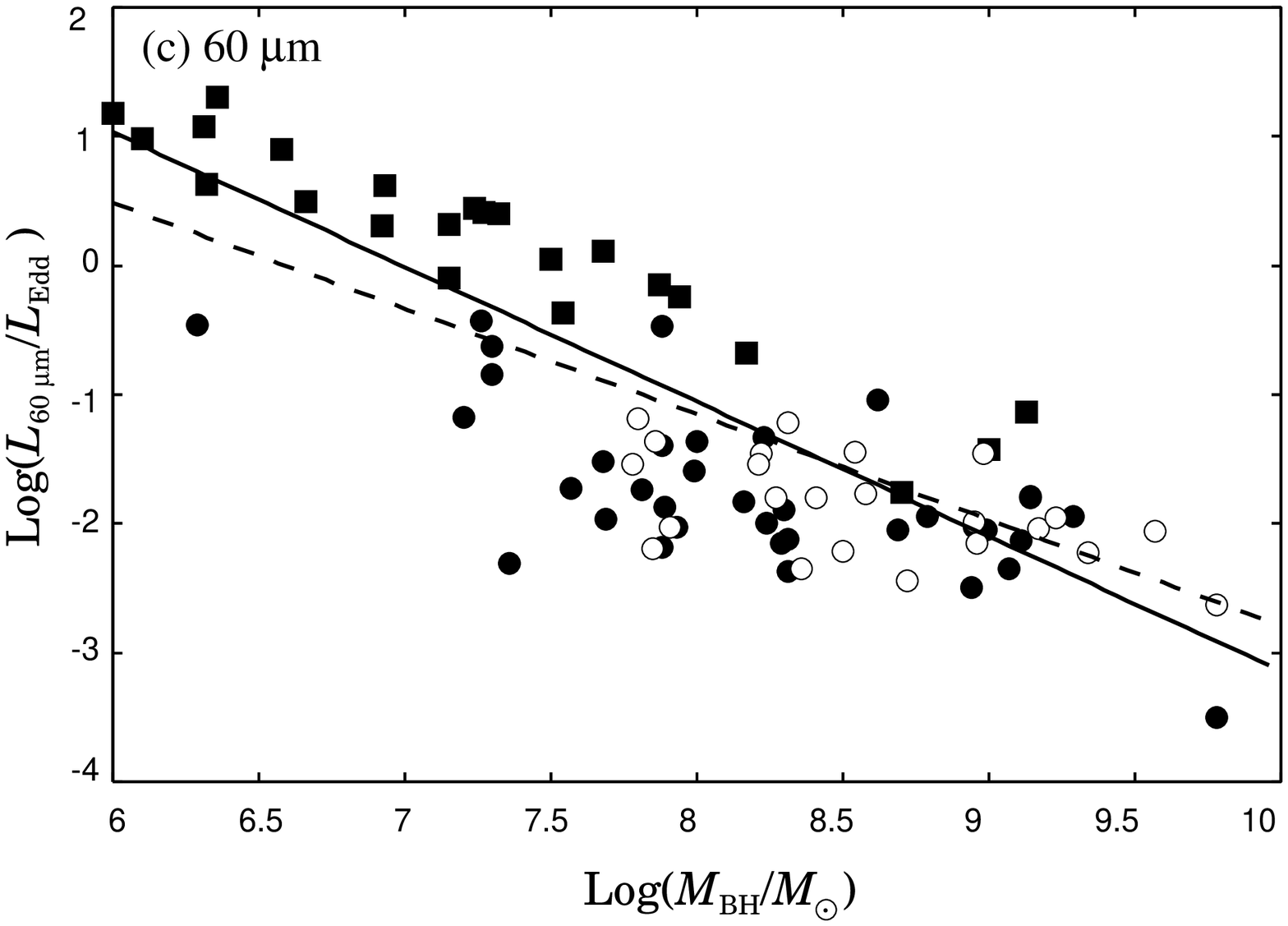}
\includegraphics[width=6cm]{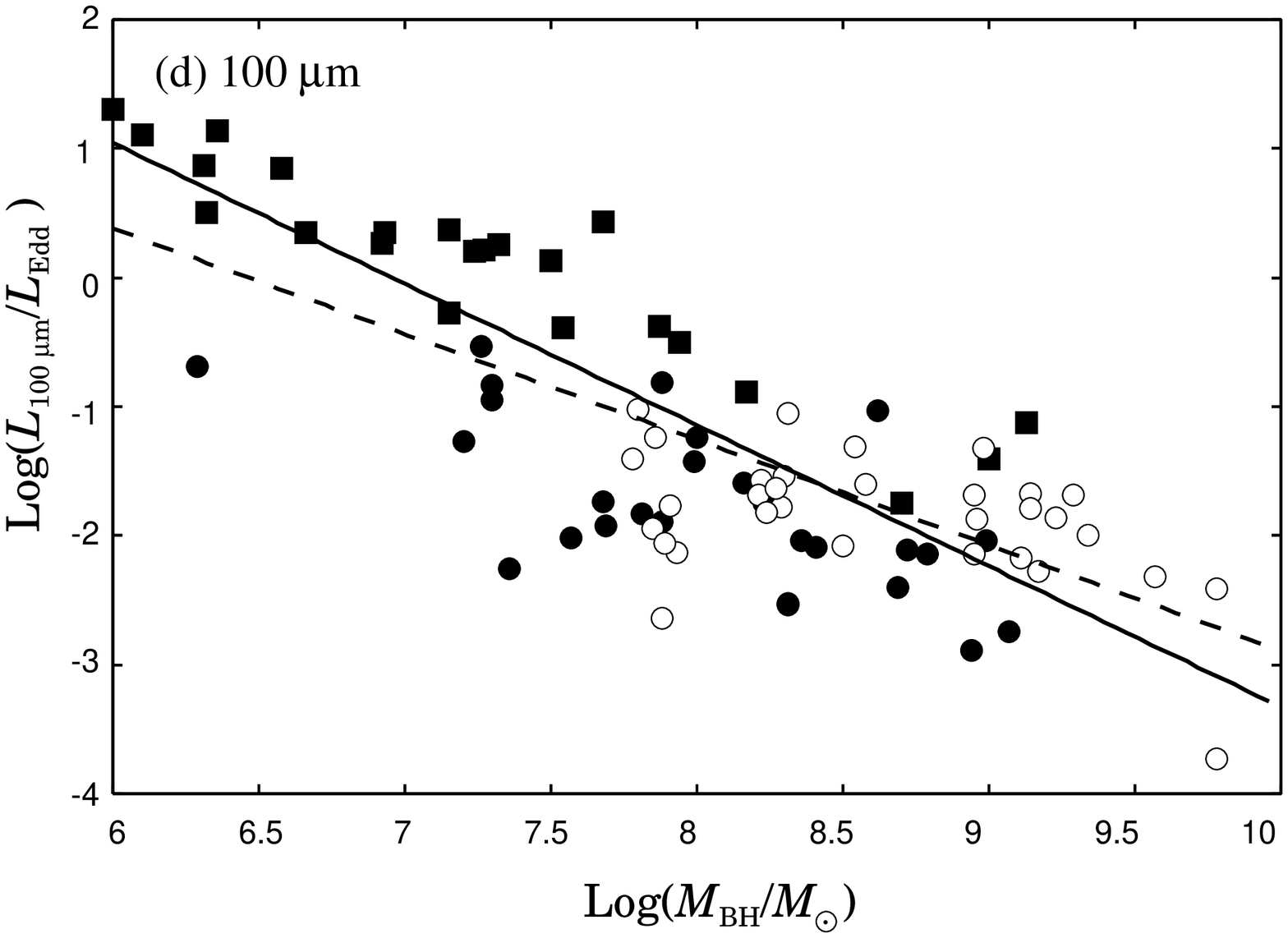}
\end{center}
\caption
{
(a) Ratio of 12-$\mu$m luminosity, defined as $\nu L_{\nu}$, 
to active galactic nuclei (AGN) Eddington luminosity ($L_{12\,\mu{\rm m}}/L_{\rm Edd}$) vs. 
black hole mass ($M_{\rm BH}$) for 23 ultraluminous infrared galaxies with type I Seyfert nuclei (squares) and 
58 PG QSOs (circles). 
Open symbols denote the upper limit.
The best-fitting relationship (solid line) 
for all targets except for the upper limit data 
was $\log(L_{12\,\mu{\rm m}}
/L_{\rm Edd})
=-0.687(\pm 0.135)\log{M_{\rm BH}}+4.98(\pm 1.0)$, 
with $\chi^{2}=0.248$. 
(b) Same as (a), but the abscissa are for the ratio of the 
25-$\mu$m luminosity to AGN Eddington luminosity, 
$L_{25\,\mu{\rm m}}/L_{\rm Edd}$. 
The best-fitting relationship (solid line) for all targets except for the upper limit data was $\log(L_{25\,\mu{\rm m}}/L_{\rm Edd})=-0.816(\pm 0.075)\log{M_{\rm BH}}+5.39(\pm 0.60)$, 
with $\chi^{2}=0.254$. 
For comparison, we overplotted the best-fitting relationship of $\log(L_{12\,\mu{\rm m}}/L_{\rm Edd})$ vs. $\log{M_{\rm BH}}$. 
(c) Same as (a), but the abscissa are for the ratio of the 
60-$\mu$m luminosity to AGN Eddington luminosity, $L_{60\,\mu{\rm m}}/L_{\rm Edd}$. 
The best-fitting relationship (solid line) for all targets except for the upper limit data was $\log(L_{60\,\mu{\rm m}}/L_{\rm Edd})=-1.04(\pm 0.099)\log{M_{\rm BH}}+7.19(\pm 0.78)$, 
with $\chi^{2}=0.46$.
(d) Same as (a), but the abscissa are for the ratio of the 
100-$\mu$m luminosity to AGN Eddington luminosity, 
$L_{100 \mu{\rm m}}/L_{\rm Edd}$. 
The best-fitting relationship (solid line) for all targets except for the upper limit data was $\log(L_{100 \mu{\rm m}}/L_{\rm Edd})=-1.09(\pm 0.115)\log{M_{\rm BH}}+7.50(\pm 0.89)$, 
with $\chi^{2}=0.49$.
}
\end{figure}

\clearpage
\begin{figure}
\plotone{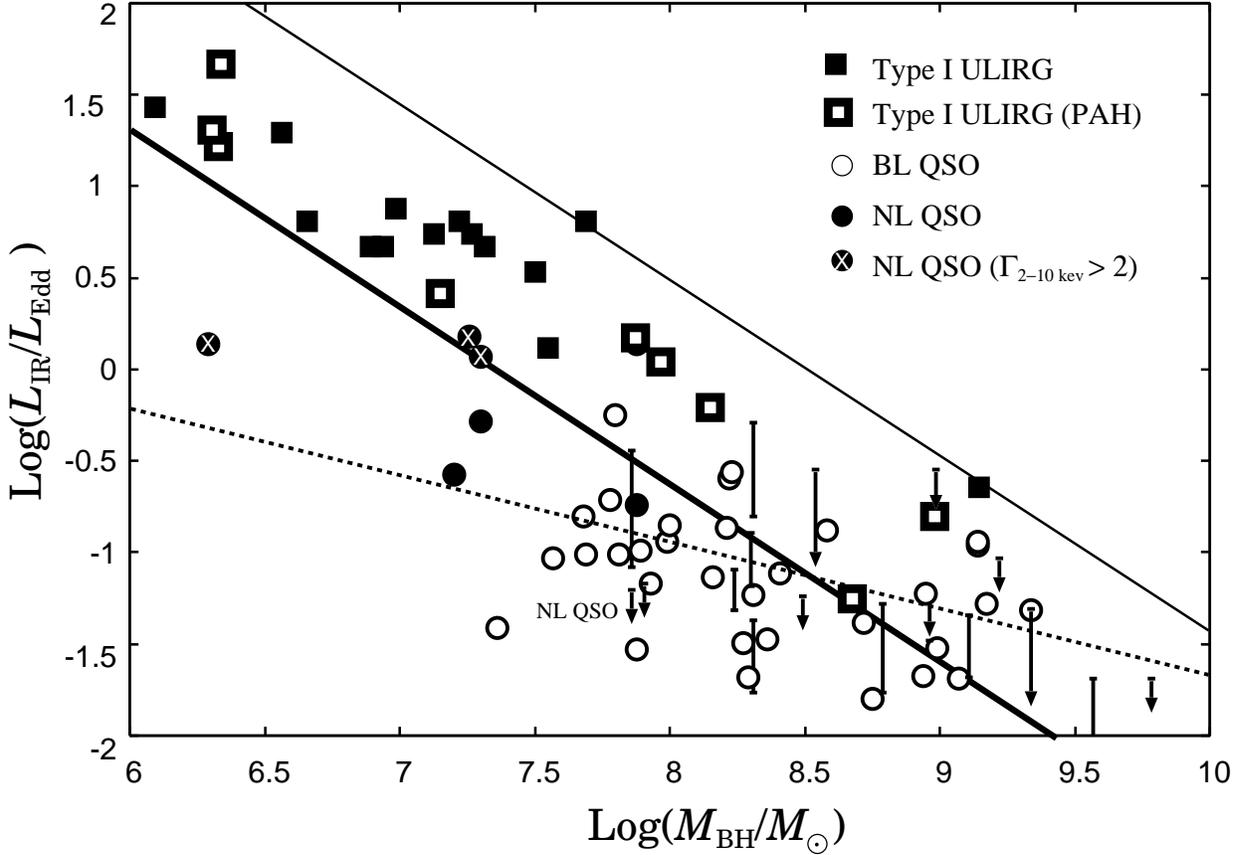}
\caption
{
Ratio of infrared to Eddington luminosity 
($L_{\rm IR}/L_{\rm Edd}$) vs. black hole mass for 23 
ultraluminous infrared galaxies with type I Seyfert nuclei (type I ULIRGs; squares) and 58 PG QSOs (circles $+$ down arrows). 
Open squares are type I ULIRGs in which polycyclic aromatic hydrocarbon
emissions were detected. 
Filled circles show narrow-line QSOs (NLQSOs), and filled circles with 
crosses denote NLQSOs with $\Gamma_{2-10\,{\rm kev}}>2$.
The thick solid line denotes the linear regression for 
all targets (type I ULIRGs and PG QSOs) except those with 
upper limit data. The best-fitting relationship for all targets was $\log(L_{\rm IR}/L_{\rm Edd})=-0.961(\pm 0.081)\log{M_{\rm BH}}+7.06(\pm 0.65)$, 
with $\chi^{2}=0.319$. 
The dotted line represents the best 
fitting relation for broad-line QSOs only, except for the upper limit data, $\log(L_{\rm IR}/L_{\rm Edd})=-0.369 (\pm 0.11)\log{M_{\rm BH}}+1.89(\pm 0.028)$, with ($\chi^{2}=0.149$). The thick and thin solid lines correspond to $L_{\rm IR}=10^{12}L_{\odot}$ and $L_{\rm IR}=10^{13}L_{\odot}$, respectively 
(see $\S 4.2$ for details). 
}
\end{figure}


\clearpage
\begin{figure}
\plotone{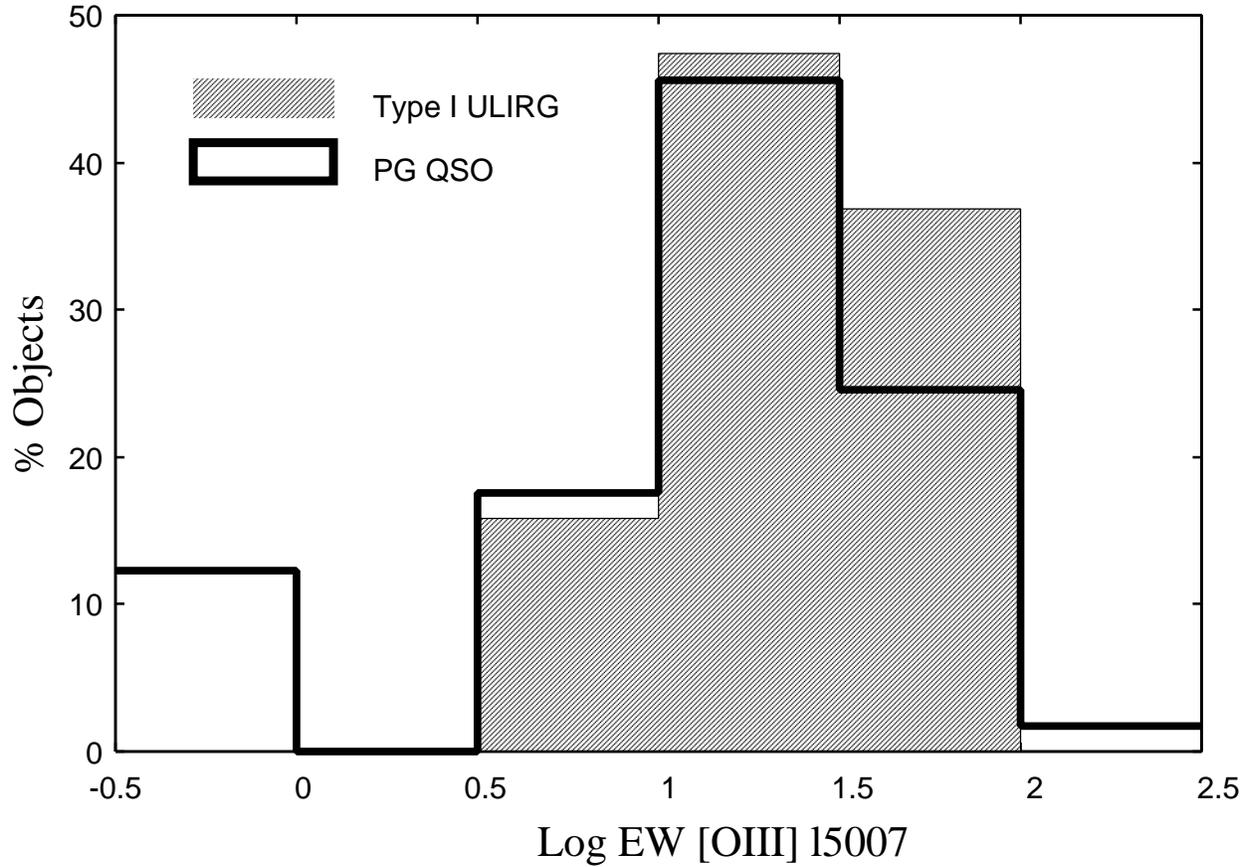}
\caption
{
Equivalent width of $[{\rm O\,III}]\lambda5007$ (rest frame) 
histogram. 
{\it Shaded bars:} The distribution of EW$[{\rm O\,III}]\lambda5007$ 
(rest frame) for ultraluminous infrared galaxies with type I Seyfert nuclei (type I ULIRGs). 
{\it Open bars:} The distribution of EW$[{\rm O\,III}]\lambda5007$ 
(rest frame) for PG QSOs. 
}
\end{figure}

\clearpage
\begin{figure}
\includegraphics[width=6cm,angle=270]{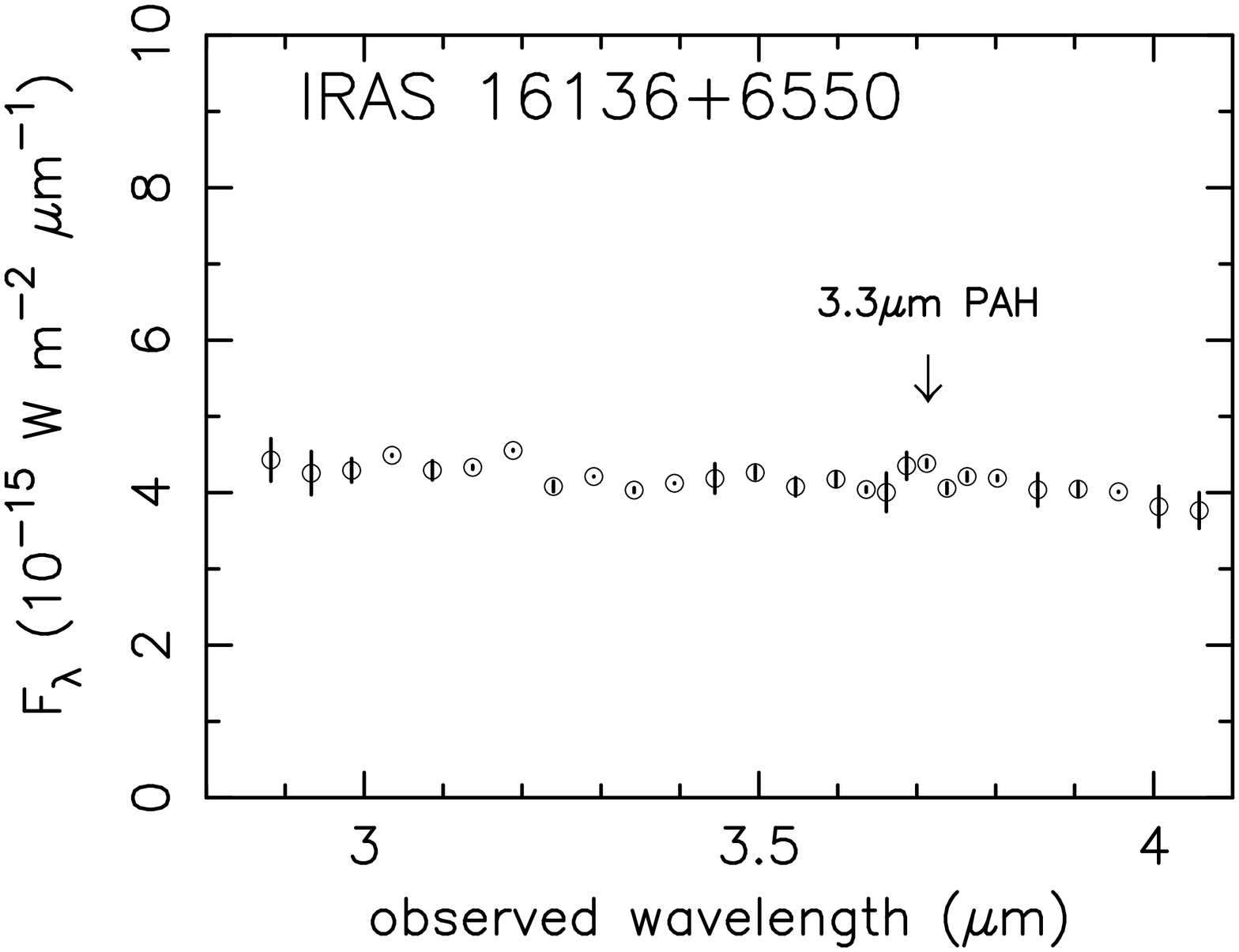}
\includegraphics[width=6cm,angle=270]{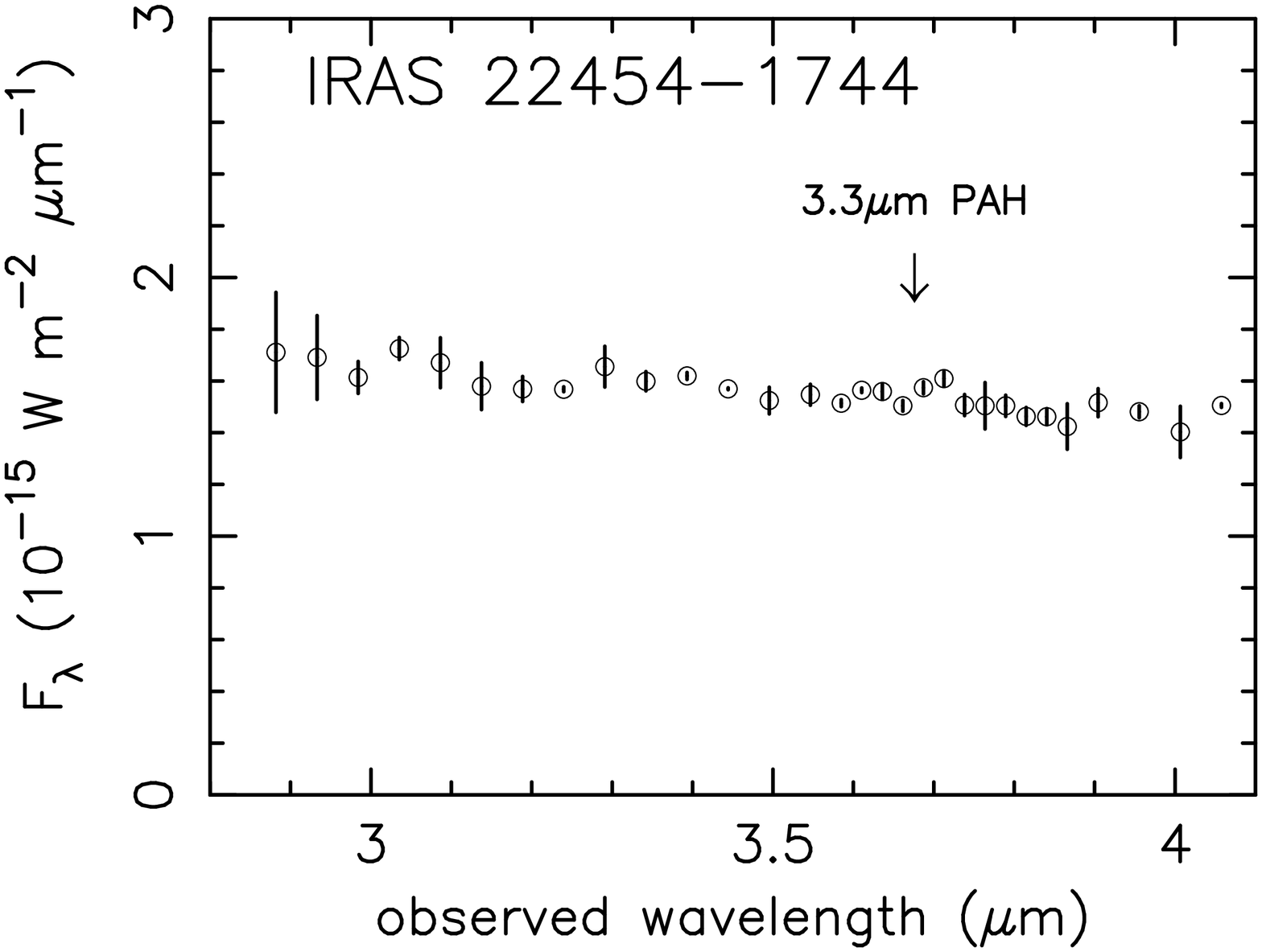}
\caption
{
Infrared $L$-band ($\lambda_{\rm obs}$ = 2.8-4.1 $\mu$m) spectra of the
two ultraluminous infrared galaxies with type I Seyfert nuclei (type I ULIRGs). 
The abscissa and ordinate are the observed wavelength in $\mu$m and
F$_{\lambda}$ in 10$^{-15}$ W m$^{-2}$ $\mu$m$^{-1}$, respectively.
The down arrows with 3.3-$\mu$m PAH (polycyclic aromatic hydrocarbon) 
indicate the expected wavelength
of the 3.3-$\mu$m PAH emission ($\lambda_{\rm rest}$ = 3.29 $\mu$m).
}
\end{figure}

\end{document}